\documentclass[fleqn,usenatbib]{mnras}

\usepackage{newtxtext,newtxmath}

\usepackage[T1]{fontenc}

\DeclareRobustCommand{\VAN}[3]{#2}
\let\VANthebibliography\thebibliography
\def\thebibliography{\DeclareRobustCommand{\VAN}[3]{##3}\VANthebibliography}

\usepackage{graphicx}	
\usepackage{amsmath}	
\usepackage{xcolor}     

\title[ML for ZTF CV Discovery]{Machine Learning Applications for Cataclysmic Variable Discovery in the ZTF Alert Stream}

\author[D. Mistry et al.]{
D. Mistry,$^{1}$\thanks{}
C. M. Copperwheat,$^{1}$
M. J. Darnley$^{1}$
and I. Olier$^{2}$
\\
$^{1}$Astrophysics Research Institute, Liverpool John Moores University, IC2, Liverpool Science Park, 146 Brownlow Hill, Liverpool, L3 5RF, UK\\
$^{2}$Data Science Research Centre, Liverpool John Moores University, James Parsons Building, 3 Byrom Street , Liverpool, L3 3AF, UK
}

\date{Accepted XXX. Received YYY; in original form ZZZ}

\pubyear{2015}

\begin{document}
\label{firstpage}
\pagerange{\pageref{firstpage}--\pageref{lastpage}}
\maketitle

\begin{abstract}
Cataclysmic variables (CV) encompass a diverse array of accreting white dwarf binary systems. Each class of CV represents a snapshot along an evolutionary journey, one with the potential to trigger a type Ia supernova event. The study of CVs offers valuable insights into binary evolution and accretion physics, with the rarest examples potentially providing the deepest insights. However, the escalating number of detected transients, coupled with our limited capacity to investigate them all, poses challenges in identifying such rarities. Machine Learning (ML) plays a pivotal role in addressing this issue by facilitating the categorisation of each detected transient into its respective transient class. Leveraging these techniques, we have developed a two-stage pipeline tailored to the ZTF transient alert stream. The first stage is an alerts filter aimed at removing non-CVs, while the latter is an ML classifier produced using XGBoost, achieving a macro average AUC score of 0.92 for distinguishing between CV classes. By utilising the Generative Topographic Mapping algorithm with classifier posterior probabilities as input, we obtain representations indicating that CV evolutionary factors play a role in classifier performance, while the associated feature maps present a potent tool for identifying the features deemed most relevant for distinguishing between classes. Implementation of the pipeline in June 2023 yielded 51 intriguing candidates that are yet to be reported as CVs or classified with further granularity. Our classifier represents a significant step in the discovery and classification of different CV classes, a domain of research still in its infancy.
\end{abstract}




\section{Introduction}

Cataclysmic variables are a class of compact binary star system in which a donor star, usually a low mass main sequence star, transfers mass via an accretion disk to a CO white dwarf (WD) via the mechanism of Roche lobe overflow \citep{RN9,RN445}. The particulars of mass transfer rate, accretion rate, donor and WD mass, orbital separation, and magnetic field strength contribute to the variety of observable phenomena that these systems display. The classification structure of CVs is based on a combination of photometric variability, X-ray characteristics, spectroscopy, and polarimetry measurements.

Thermal and viscous instabilities in the accretion disk, described by various incarnations of the disk instability model \citep{RN496,RN331,RN395}, cause semi-regular brightening events each referred to as a dwarf nova outburst. Systems that undergo dwarf nova outbursts are named dwarf novae. Outburst amplitudes typically lie in the 2-5 magnitude range usually lasting between a few days to a fortnight, recurring on timescales of days to months; these attributes are specific to a given system. The dwarf nova class can be further subdivided into the U Geminorum (U Gem), Z Camelopardalis (Z Cam), and SU Ursae Majoris (SU UMa) subclasses. Standstills (periods of constant brightness) and superoutbursts (dwarf novae outbursts of greater amplitude and duration), respectively, distinguish the Z Cam and SU UMa subclasses from one another and U Gem, which display only 'normal' outbursts (e.g., \cite{RN367,RN491}. Extremes in the recurrence times of superoutbursts (supercycles) facilitate two major SU UMa subclasses, the ER Ursae Majoris type \citep{RN529}, distinguished by extremely short supercycles with rapid fire normal outbursts in between, and WZ Sagittae systems that appear to display no normal outbursts, only superoutbursts, with supercycle lengths of order years (e.g., \citealt{RN530}).

Stable (hot and viscous) accretion disks give rise to systems with almost constant brightness, referred to as nova-likes. Some nova-like stars undergo periods where mass transferred from the donor is either diminished or even completely suppressed. Consequently, a drop in brightness of 3–6 mag in the optical occurs. These systems are referred to as VY Sculptoris stars \citep{RN383,RN397} - a nova-like subtype. Whilst, there exist four nova-like subtypes, three distinguishable spectroscopically, VY Scl is the only one that can be identified photometrically.

Novae are modelled as thermonuclear runaway events within the accreted layer of hydrogen on the WD surface (e.g., \citealt{RN75,RN428,RN167,RN62}, they produce a sudden high amplitude (8-15 magnitudes typically) increase in optical brightness with a long duration decline (weeks to years). Recurrence times are largely dependent on the WD mass and donor mass transfer rate. Recurrent novae (RNe) have been observed to undergo more than one nova eruption, with recurrence times below 100 years, while for classical novae (only one eruption observed) this is likely to extend up to 100,000 years.

Where the WD is strongly magnetic, with magnetic fields of $B > 10 MG$, the formation of an accretion disk is inhibited, instead the mass transfer stream from the donor is directed out of the orbital plane and funneled by the magnetic field lines directly onto one or both of the WD's magnetic poles. Referred to as polars, or AM Herculis stars \citep{RN307,RN430}, the WD rotates synchronously with the orbital period causing the accretion flow to always interact with same field lines. Their photometric variability is a consequence of the complex interplay between intrinsic and extrinsic sources of variability; high and low states of brightness (as a consequence of fluctuations in the rate of mass transferred by the donor) and orbital period modulations of few tenths of a magnitude, due to obscuration of the accretion flow or spot behind the limb of the WD, contribute to the superposition of both short (hours) and long (weeks to months) timescale photometric variability.

Intermediate polars (or DQ Herculis stars) represent the intermediary between polars and non-magnetic CVs with magnetic field strengths of between 1 and 10 MG \citep{RN313,RN514}. A partial accretion disk may form, though disruption occurs closer to the WD causing magnetically controlled accretion at smaller radii. Photometrically, they can display modulations due to the WD spin period (non-synchronous rotation) and the sideband period between the spin and orbital periods. Possibly also present are low amplitude dwarf nova outbursts due to the truncated nature of the accretion disk, and high and low states in brightness (e.g., \citealt{RN565}).

\begin{figure*}
    \centering
    \includegraphics[width=0.95\textwidth]{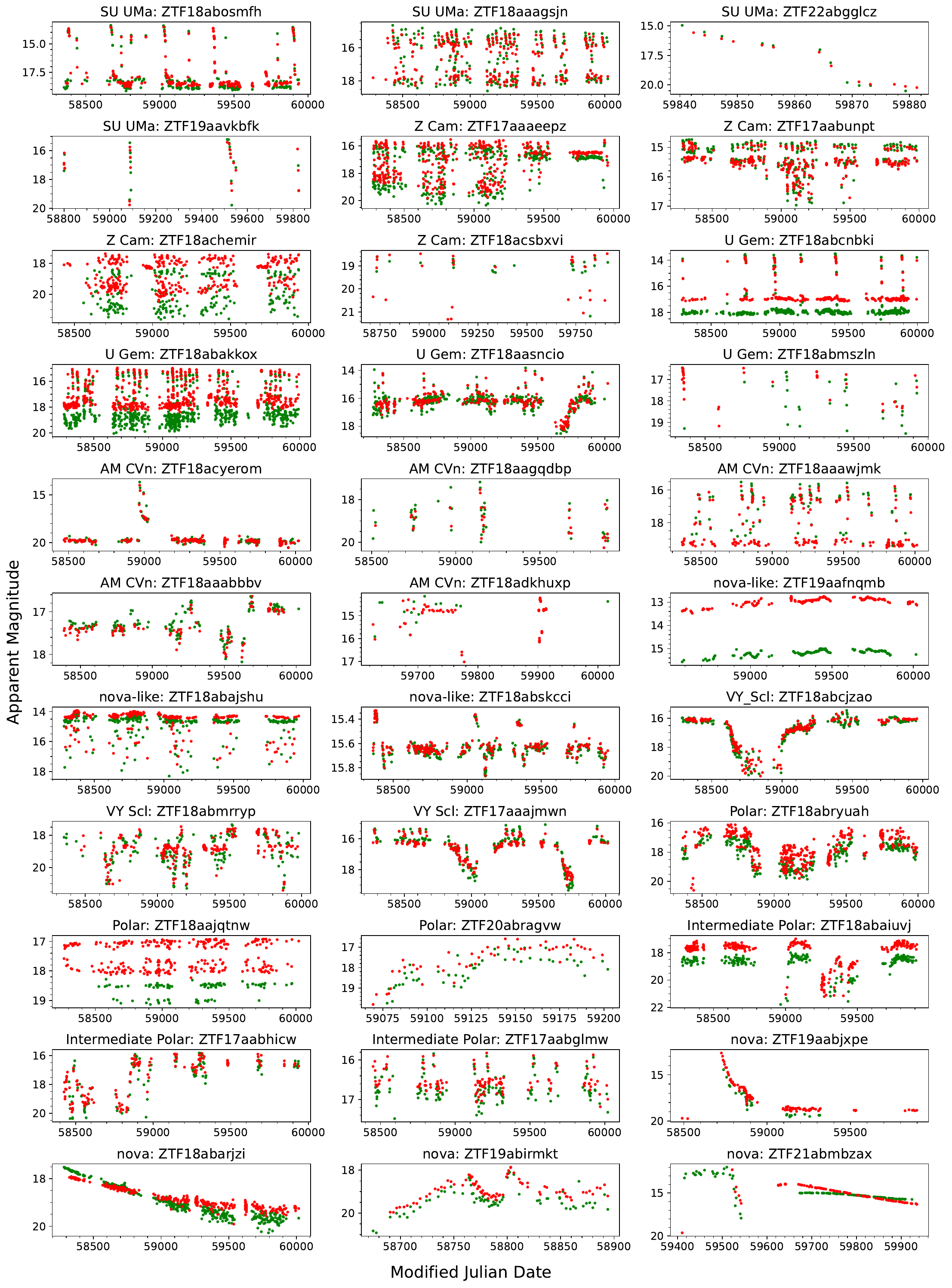}
    \caption{Example light curves of each CV class. Green and red points indicate g and r band observations respectively.}
    \label{fig:class_lightcurves}
\end{figure*}

The AM Canum Venaticorum stars \citep{RN70,RN480} are rare and ultra-short period (5-65 minutes) binaries where the donor is believed to be either another degenerate or a semi-degenerate star composed mostly of helium. They are characterised by their blue colour, due to the WD dominating the flux contribution over an extremely low mass donor (within Gaia DR3 \citealt{RN545} the BP-RP colour is typically less than 0.6); strong helium emission and absence of hydrogen within their spectra. Photometrically, their variability is dependent on their orbital period, those with period between 22 and 51 minutes tend to display the outbursting characteristics of their hydrogen rich counterparts though with lower duration and amplitude \citep{RN482}. The diversity of ZTF light curves for each class of CV is shown in Figure \ref{fig:class_lightcurves}.

Wide field, high cadence and panchromatic surveys such as the Sloan Digital Sky Survey (SDSS; \citealt{RN552}), the (intermediate) Palomar Transient Factory (iPTF; \citealt{RN554}), the Dark Energy Survey (DES; \citealt{RN555}), and new ongoing surveys such as the Zwicky Transient Facility (ZTF; \citealt{RN68}), the Asteroid Terrestrial-impact Last Alert System' (ATLAS; \citealt{RN259}), and the Gravitational-wave Optical Transient Observer (GOTO; \citealt{RN556}) have dramatically improved our ability to discover these objects. These discoveries have helped fill gaps in our current knowledge, for example, constantly evolving models are being developed attempting to explain the diversity of dwarf nova outbursts based on the disk instability model \citep{RN316,RN531}; \cite{RN521} was able to construct semi-empirical models for the evolution of CVs based on donor star masses and radii; the discovery of a genuine standstill in the AM CVn CR Boo helps to support the viewpoint of AM CVns being the helium rich analogue of hydrogen rich CVs \citep{RN511}; and the larger sample size of all CVs facilitates a disentanglement of CV subtypes within a H-R diagram \citep{RN497}. However, discoveries also uncover new gaps too, such as the detection of pulsed X-rays in two AM CVns that not only raises the question of magnetically controlled accretion in AM CVns but leads to implications for their evolutionary timescales \citep{RN527}. To accelerate our current understanding of CVs and in turn accretion mechanisms in transients such as X-ray binaries and active galactic nuclei (AGN), we require a greater sample size of CV members, especially the more elusive subclasses.

The identification of new and rare/unique CV candidates from survey data is becoming an ever more difficult task due to the challenge of efficiently finding them amongst the large numbers of sources exhibiting significant variability that are detected every night. Sources responsible include, but are not limited to, supernovae, variables stars, AGN, tidal disruption events, and Solar System objects that include asteroids, as well as artifacts (bogus alerts). In the case of ZTF, transient alert rates can exceed a million per night \citep{RN376}, and this rate is set to be dwarfed by the Rubin Observatory \citep{RN60}. As a further side effect, facilities devoted to the follow-up of transient events are not enough in number to investigate them all, therefore time on such facilities is in short supply. Since the majority of genuine astrophysical sources may serve only to reaffirm our current understanding of the transient classes to which they belong, follow-up time will be reserved for the minority, those that present a challenge to or help further our understanding.

Machine learning (ML) is widely acknowledged as a powerful set of techniques ideally suited to address these challenges, with applications to source classification. For example , the classification of ZTF alerts by \cite{RN421}, CRTS light curves by \cite{RN65} and the recent utilisation of ML in the separation of Gaia transients into over 25 different classes \citep{RN515}. Specific focus on automated identification of CVs and their subtypes is an active, though underdeveloped, field of research. Examples so far include the 497 CVs uncovered from ZTF alerts by applying simple colour, amplitude and variability timescale filters \citep{RN69, RN408}; an extension of this filter approach by \cite{RN477}, employing Gaia and PanSTARRS colours to identify nine outbursting AM CVns within ZTF alerts; and application of ML to identify CVs within Gaia Science Alerts \citep{RN532}. 

Here are presented details of our development and application of an automated ML pipeline aimed at identifying the various classes of CVs from the ZTF alert stream via the Lasair alerts broker \citep{RN377}. We start by explaining the initial alerts filtering using Lasair (Section \ref{sec:Lasair_filter}) before moving on to describing the construction of our dataset upon which a ML classifier is generated (Sections \ref{sec:source_list} - \ref{sec:features}). Sections \ref{sec: train_test_split} - \ref{sec:performance_metrics} describe the ML techniques adopted and algorithms tested. The results of our efforts to generate a suitable ML CV classifier for our pipeline are presented in Section \ref{sec:results} along with its initial outcomes based on implementation. The discussion of our results (Section \ref{sec:discussion}) will be given in the context of light curve profiles and the underlying physical properties of the CV subtypes.

\section{Method}

\subsection{Alerts filter}\label{sec:Lasair_filter}

Alert streams from ZTF are ingested by alerts brokers such as Lasair \citep{RN377} and Alerce \citep{RN421}. They provide real-time alerts access, as well as science, difference and reference image cutouts, light curves of the associated ZTF object, contextual information, statistics derived from source photometry, and the ability to cross-match events with catalogued sources. Brokers provide the ability to filter alerts based on the above in order to focus on those that are most relevant to their science goals. Our pipeline experiments with Lasair's cross-matching and filtering services to focus on objects that lie within the typical parameter space of CVs as a first stage before implementation of our ML classifier.

\begin{figure*}
    \centering
    \includegraphics[width=\textwidth]{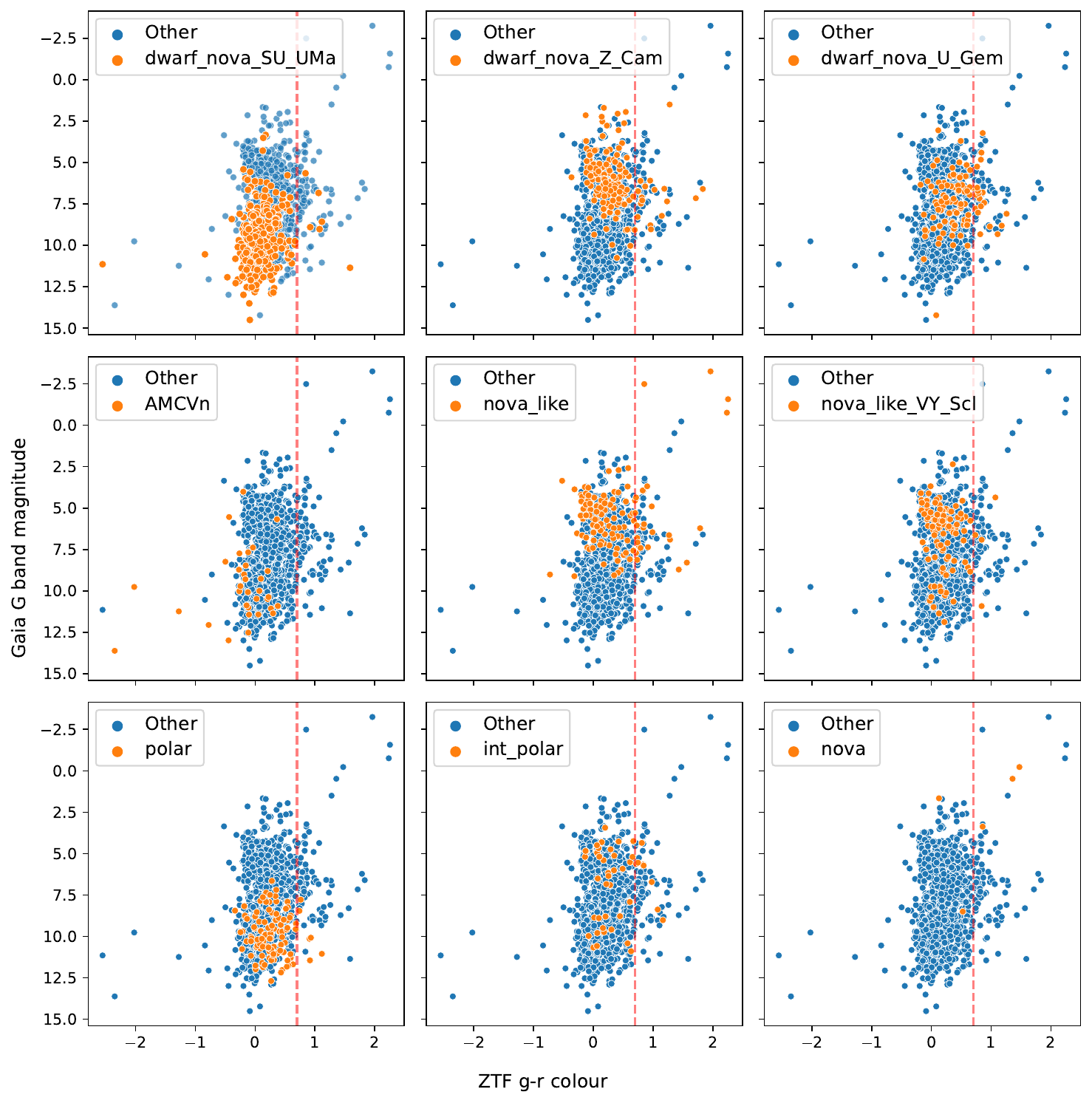}
    \caption{Colour magnitude diagrams using Gaia G band absolute magnitude and the colour derived from the ZTF g and r bands. The dashed red line in each plot denotes the ZTF g-r colour threshold of 0.7. Orange points in each subplot denote examples of a particular CV class, while the blue points represent examples belonging to the remaining classes (labelled ''other'').}
    \label{fig:cmd_gminusr}
\end{figure*}

To remove non-CV catalogued sources, the Sherlock classification software \citep{RN543}, implemented by Lasair for cross-matching, is examined. Sherlock uses a model, generated by a boosted decision tree algorithm, that mines a database of historical and on-going astronomical survey data to predict the nature of the object based on the resulting crossmatches. The database include datasets from all-sky surveys as well as more source specific catalogues such as the Million Quasars Catalog \citep{RN544}, Downes Catalog of CVs \citep{RN516}, and the Ritter Cataclysmic Binaries Catalog v7.24 \citep{RN443}. Sherlock assigns the label Variable Star (VS), Cataclysmic Variable (CV), Active Galactic Nuclei (AGN), or nuclear transient (NT) should the transient be located within the synonym radius (1.5") of a catalogued point source or, in the case of a NT, the core of a resolved galaxy; a supernova (SN) if not classified as a NT but is found close enough to a resolved galaxy to be deemed physically associated; a Bright Star (BS) if the transient is not matched against the synonym radius of a star but is associated within the magnitude-dependent association radius; Orphan if the transient fails to be matched with a catalogue source; or Unclear otherwise. In an effort to limit alerts of non-CVs we experimented with the use of Sherlock, and catalogue cross-matching. The remaining sources are then subject to colour and magnitude change cuts akin to those described in \cite{RN69,RN408}. In those works, the ZTF alert stream filtering involved looking for point sources with g-r colour < 0.6 and a magnitude change $\Delta m>=2$ within a timescale of 2 days in the g band. This resulted in a total of 701 known or candidate CVs over two years of its implementation that typically displayed dwarf nova outbursts and changes in accretion state. We relaxed these constraints with respect to \cite{RN69,RN408} to maximise the number of targets for classification. In performing a cut based on colour, attempts were made to account for several factors: differences in the sampling between the g and r band; sampling differences between outburst activity and quiescence; and the tendency of CVs to have bluer colours during outbursting phases than during quiescence (a consequence of the enhanced accretion and increased temperature of the disk during outburst). Therefore, for each source, the colour for each night of observation was extracted (where calculable); the mean and median averages of these were recorded along with the colour at maximum and minimum brightness. The constraint of <=0.7 for each of these quantities, as well as for the overall mean colour (calculated without the epochal requirement) was utilised. Figure \ref{fig:cmd_gminusr} shows that a significant fraction of CVs will be recovered at or below the epochal mean g-r of 0.7. This constraint is flexible, based on the type of CV we may wish to focus our attention on. Constraints placed on magnitude change, $\Delta m$, involved experimenting with various thresholds. A higher $\Delta m$ yielded sources with more rapid variability, e.g., Z Cam systems, while lower values increased the contribution of sources akin to nova-likes. Given that alerting sources that the filter outputs are entered into a ML classifier to distinguish these variability differences, foregoing a $\Delta m$ constraint is the approach adopted.

\subsection{Source List} \label{sec:source_list}

The light curves and associated metadata (see the following subsection) of the sources remaining after the Lasair filter are used as input for a ML based CV subclass classifier. The classifier is trained on the ZTF g and r band light curves of catalogued CVs whose subtypes have been ascertained along with associated Gaia Data Release 3 data \citep{RN545} where available. This section describes the nature of the data set for training and testing of candidate classifiers.

To construct a dataset we consulted the American Association of Variable Star Observers Variable Star Index (VSX\footnote{https://www.aavso.org/vsx/index.php}) which is a continuously updated repository of transient sources. Confirmed CVs from archival resources such as The Catalogue and Atlas of Cataclysmic Variables \footnote{https://heasarc.gsfc.nasa.gov/W3Browse/all/cvcat.html} \citep{RN516}, and the catalogue of cataclysmic binaries, low-mass X-ray binaries and related objects\footnote{https://heasarc.gsfc.nasa.gov/W3Browse/all/rittercv.html} \citep{RN443} are contained within the repository, as are more recent discoveries detailed in literature (e.g., \citealt{RN403,RN69,RN465}). Each repository source has a dedicated page where further information can be found such as their designated names in other surveys, references to literature for that source, orbital periods, and more. The labelling procedure conducted by VSX involves constant review and revision of metadata, with citations for any new details and rationales behind changes fully documented. VSX contained a list of over 15,300 targets classified as CV, of which 5683 were successfully cross matched with ZTF alerts objects. We supplemented this list with novae catalogued in the Bright Transient Survey\footnote{https://sites.astro.caltech.edu/ztf/bts/bts.php} (BTS) and not in the AAVSO list. This constituted an extra 28 sources making a total of 5708 CVs. The vast majority (4822) were of the dwarf nova subclass. Since we aim for a more granular classification than that, we refined our sample further to only include dwarf nova examples with further subdivision into the U Gem, Z Cam, and SU UMa subtypes. This left us with a dataset of 1568 samples.

\subsection{Light curves}

The light curves themselves are generated from observations with the 47 square-degree camera mounted on the Samuel Oschin Telescope at Palomar Observatory in California \citep{RN557}. For a 30 second exposure the median 5$\sigma$ limiting magnitude is 20.8 in the g band and 20.6 in the r band. The observing strategy involves three surveys, the g and r band data for two of which are available publicly. The Northern sky survey is a three-day cadence survey of all fields north of -31$\deg$, while the Galactic plane survey observes daily within 7$\deg$ of the Galactic plane. For both surveys, each night a field is observed, it is observed twice, once for each of the g and r bands, and at least 30 minutes between visits. With these cadences, superoutbursts, whose durations range from a few days to several weeks, are well sampled, as are nova eruptions, high and low states of brightness, and standstills. The g and r band also provide colour information, a further tool for class separation.

Light curves of cross matched sources were downloaded from Lasair. Brightness values are given in difference magnitudes, this is the magnitude derived from the positive difference between the flux in the reference image and that in the science image. Where a source contains data points below the reference flux, the difference magnitude light curve profile may deviate from that one would expect for its transient class. Subsequently, these difference fluxes were converted to apparent magnitudes where possible. The formulae used to convert from difference magnitudes to apparent magnitudes and associated errors are given by:

\begin{equation}
    m_{\rm corr} = -2.5 \, \log_{10}(10^{-0.4 \,m_{\rm ref}}+sgn \, 10^{-0.4\, m_{\rm diff}})
\end{equation}

\begin{equation}
    \delta m_{\rm corr} = \frac{(10^{-0.8\,m_{\rm diff}}\,\delta m^{2}_{\rm diff})^{0.5}}{10^{-0.4 \, m_{\rm ref}}+sgn\,10^{-0.4\, m_{\rm diff}}}
\end{equation}

where we simply convert the difference $m_{\rm diff}$ and reference $m_{\rm ref}$ magnitudes to fluxes, sum them considering the sign of the alert ($sgn$) and convert the results back to magnitude $m_{\rm corr}$. Simple error propagation gives the error $\delta \, m_{\rm corr}$.

To be included into our dataset, two main vetting procedures were followed. The first was to verify the label by checking the references associated with the source. This was easier for the less prevalent classes such as the magnetic systems and AM CVns, where membership can only be verified by means beyond photometry (e.g., spectroscopy, and pulsed X-ray detection), and for dwarf novae further subdivided into the SU UMa and Z Cam classes. For U Gem dwarf novae and those dwarf novae not divided into subclasses, references to literature were less readily available. A second vetting procedure involved inspection of the light curves themselves, where missclassifications were identified and their appropriateness for dataset inclusion could be assessed. In assessing their suitability for inclusion we considered whether phenomena characteristic to a given transient type (e.g., standstills or nova eruption) were present, the number of datapoints, and whether colour information may be derived. One must be careful to omit examples based on the number of data points, as a limited number may be representative of sources only visible during brightening events. With this consideration in mind, a minimum threshold of at least four points in at least one filter was set.

Example ZTF light curves for each of the classes defined in the following section are given in Figure \ref{fig:class_lightcurves}. Aside from the usual observing gaps due to the time of year, the limiting magnitude of the telescope in combination with the brightness of the source results in a variety of observational timespans - objects below the limiting magnitude in quiescence may briefly rise into view during episodes of activity, e.g., ZTF22abgglcz and ZTF19aavkbfk. Outbursts of different cycle lengths (time between successive outbursts) are clearly evident for dwarf novae, as are superoutbursts (e.g., ZTF18abosmfh). Evident also are standstills (e.g., ZTF17aaaeepz), long term changes (high and low brightness states) due to changes in mass transfer rate (e.g., ZTF18aasncio, ZTF18abcjzao, and ZTF18abryuah), and the various outburst profiles of nova eruptions.

\subsection{Classification structure}

With our task firmly routed in distinguishing between the different types of CV, we settled on a nine class classification structure that separated the dwarf nova class into their three main subtypes, SU UMa, Z Cam, and U Gem; distinguished between nova-likes and nova-likes containing the VY Scl characteristic (see introduction); separated the magnetic CVs into their polar and intermediate polar subclasses; with novae and AM CVns making up the remainder. The structure is motivated by the desire for a model that classifies to the highest level of class granularity (to group examples by their most unique traits) while at the same time balancing this desire with the requirement of enough examples to represent the class. This unfortunately inhibits our ability to separate the WZ Sge and ER UMa systems from their parent class (SU UMa), and separate novae by their various light curve profiles.

Table \ref{tab:class numbers} shows the number of examples per CV class following our vetting procedures. The list is understandably heavily biased towards dwarf novae due to their ubiquity within the CV population.

\begin{table} \label{table:class numbers}
    \centering
    \begin{tabular}{c|c|c}
        Class/subclass & Number of targets\\
        \hline
        SU Ursae Majoris & 630\\
        Z Camelopardalis & 174\\
        U Geminorum & 116\\
        nova-like VY Sculptoris & 120\\
        nova-like non VY Sculptoris & 123\\
        nova & 46\\
        polar & 114\\
        intermediate polar & 49\\
        AM Canum Venaticorum & 46\\
    \hline
    \end{tabular}
    \caption{Number of targets per CV class within the dataset.}
    \label{tab:class numbers}
\end{table}

\subsection{Features}\label{sec:features}

\subsubsection{ZTF Light curve derived features}

To distinguish between the classes of CV, statistical, percentile and periodicity based features were extracted from the g and r band source light curves. The suite of features provided by the feATURE eXTRACTOR FOR tIME sERIES (\textit{feets}) python package \citep{RN451} is comprehensive enough to describe the vast majority of variability characteristics present within our light curves. We therefore make use of them with the addition of several features of our own that are more specifically geared towards CV variability. Non-outbursting systems such as nova-likes and polars are generally well characterised by the \textit{feets} feature set. The diversity of outbursting systems however, are less well characterised after baseline models revealed the confusion between classes exhibiting such behaviour.

As described in \cite{RN508}, the typical observing cadence, sampling consistency (affected by weather), limiting magnitude and the number of filters that a survey operates under governs our ability to visually recognise and extract features that accurately describe the different types of variability displayed by dwarf nova exhibiting systems. Sub-optimal conditions related to the above inhibit the usefulness of the features extracted. Given the level of classification granularity desired in this work, we developed several simple features that may recognise the presence of phenomena such as superoutbursts, standstills, and their properties.

The \textit{find\_peaks} function from the \textit{scipy} python package locates signal peaks (outbursts peaks in our case) by simply comparing neighbouring brightness values. Not all peaks are identifiable due to undersampled outburst and quiescent phases, and intricacies of the function, though enough useful information is present to obtain the following: an outburst amplitude based on the peak with the largest such value; and rise and decline rates based on the minimum time between outburst peaks and their bases. These features were evaluated for specific outburst amplitude ranges. Recurrence rates are best described by the frequency at which the maximum power of the Lomb Scargle periodogram of the light curve occurs. The Lomb Scargle method will output a value even if outbursts or strong periodic signals are not observed. The ratio of the maximum power to the mean power is therefore used to distinguish strong from weak periodic signals. With respect to standstills, obvious instances can be characterised by utilising a rolling standard deviation window. Sources with standstills will have windows with high standard deviation values during outbursting periods and low values during standstills. A high ratio of the maximum of the former to the minimum of the latter can detect this dichotomy. This dichotomy however, is also present in outbursting systems with well defined quiescent phases (without standstills). One is separated from the other by including the mean brightness level of the window with the minimum standard deviation. A brightness level appreciably higher than the minimum brightness aims to provide the distinction.

Colour is a useful separator of different CV subtypes. In addition to the g-r colour calculated from the average brightness in each filter, we derive the colour for each night where both a g and r band observation was recorded. We include the mean and median of these as features to mitigate the skewing of colour values due to sampling differences between the bands during outburst and quiescence phases. Furthermore we include the colour at maximum and minimum brightness to account for bluer colours during outbursting phases. All light curve derived features are given in Tables \ref{tab:lc_features} and \ref{tab:additional_features}. 

\begin{table*} 
    \caption{Features extracted from each of the g and r band light curves. Listed are those available from the {\fontfamily{qcr}\selectfont feets} package, where for each a more detailed explanation is provided at \url{https://feets.readthedocs.io/en/latest/tutorial.html}. \label{tab:lc_features}}
    \begin{tabular}{p{0.25\textwidth}p{0.70\textwidth}}
        \hline
        Feature & Description\\
        \hline
        \textit{Amplitude} & Half of the difference between the median of the maximum 5\% and the median of the minimum 5\% magnitudes\\
        \textit{AndersonDarling} & The Anderson-Darling test is a statistical test of whether a given sample of data is drawn from a given probability distribution (normal distribution)\\
        \textit{Autocor\_length} & Cross-correlation of a signal with itself. Informally, described as the similarity between observations as a function of the time lag between them, useful for finding repeating patterns. Autocorrelation returns a vector, the feature returns the vector length for values less than $e^{-1}$.\\
        \textit{Beyond1Std} & Percentage of points beyond one standard deviation from the weighted mean (weighted by the square of the inverse error).\\
        \textit{CAR\_mean} & The mean parameter used to model irregularly sampled time series with the continuous time auto regressive model \citep{RN549}.\\
        \textit{CAR\_sigma} & The variability parameter used to model irregularly sampled time series with the continuous time auto regressive model.\\
        \textit{CAR\_tau} & The tau parameter used to model irregularly sampled time series with the continuous time auto regressive model. Interpreted as the variability amplitude of the light curve.\\
        \textit{Con} & The number of three consecutive data points that are brighter or fainter than $2 \sigma$ and normalised the number by N-2.\\
        \textit{Eta\_e} ($\eta^e$) & Variability index $\eta$ is the ratio of the mean of the square of successive differences to the variance of data points.\\
        \textit{FluxPercentileRatioMid\textbf{X}} & Ratio of centred flux percentile ranges. If $F_{5,95}$ is the difference between the 95th and 5th percentile of ordered magnitudes, then \textit{FluxPercentileRatioMid\textbf{X}} = ${F_{40,60}}/{F_{5,95}}$, ${F_{32.5,67.5}}/{F_{5,95}}$, ${F_{25,75}}/{F_{5,95}}$, ${F_{17.5,82.5}}/{F_{5,95}}$, and ${F_{10,90}}/{F_{5,95}}$, for \textbf{X} = 20, 35, 50, 65, and 80 respectively.\\
        \textit{Freq\textbf{i}\_harmonics\_amplitude\_j} & Amplitude of the jth harmonic of the \textbf{i}th frequency component of the Lomb Scargle Periodogram\\
        \textit{Freq\textbf{i}\_harmonics\_rel\_phase\_\textbf{i}} & The phase corresponding to \textit{Freq\textbf{i}\_harmonics\_amplitude\_j} relative to the phase of the first frequency component.\\
        \textit{Gskew} & Median-of-magnitudes based measure of the skew\\
        \textit{LinearTrend} & Slope of a linear fit to the light-curve\\
        \textit{MaxSlope} & Maximum absolute magnitude slope between two consecutive observations\\
        \textit{Mean} & Mean magnitude\\
        \textit{Meanvariance} & Ratio of the standard deviation to the mean magnitude\\
        \textit{MedianAbsDev} & Median absolute deviation of magnitude\\
        \textit{MedianBRP} & Median Buffer Range Percentage; Fraction (<= 1) of photometric points within amplitude/10 of the median magnitude.\\
        \textit{PairSlopeTrend} & Considering the last 30 (time-sorted) measurements of source magnitude, the fraction of increasing first differences minus the fraction of decreasing first differences\\
        \textit{PercentAmplitude} & Largest percentage difference between either the max or min magnitude and the median\\
        \textit{PercentDifferenceFluxPercentile} & Ratio of the difference between the 95th and 5th percentile of ordered magnitudes, $F_{5,95}$, over the median magnitude.\\
        \textit{PeriodLS} & Period corresponding to frequency of maximum power in the Lomb Scargle Periodogram\\
        \textit{Period\_fit} & The false alarm probability of the largest Lomb Scargle periodogram value.\\
        \textit{Psi\_CS} & \textit{RCS} applied to the phase-folded light curve (generated using the period estimated from the Lomb-Scargle method).\\
        \textit{Psi\_eta} & $\eta^e$ index calculated from the phase-folded light curve\\
        \textit{Q31} & Difference between the third and first quartile of the light curve magnitudes\\
        \textit{Rcs} & Range of a cumulative sum (R\textsubscript{CS}) of the light curve. Defined as: R\textsubscript{CS} = max(S) - min(S), where $S=\frac{1}{N\sigma}\sum^{l}_{i=1} (m_i-\Bar{m})$. $N$ represents the number of points, with $i=1,2,...,N$.\\
        \textit{Skew} & Skewness of the magnitudes\\
        \textit{SlottedA\_length} & Slotted autocorrelation length - same as \textit{Autocor\_length} except that time lags are defined as intervals or slots instead of single values\\
        \textit{SmallKurtosis} & Small sample kurtosis of the magnitudes.\\
        \textit{Std\_g} & Standard deviation of magnitudes.\\
        \textit{StetsonK} & Robust measure of the kurtosis \citep{RN550}.\\
        \textit{StetsonK\_AC} & Variability index derived based on the autocorrelation function of each lightcurve \citep{RN550}.\\
        \textit{StructureFunction\_index\_21} & \\
        \textit{Q31\_colour} & \textit{Q31} applied to the difference in the g and r band magnitudes.\\
        \textit{StetsonJ} & A robust version of the Welch/Stetson variability index I \citep{RN550} describing the synchronous variability of different bands.\\
        \textit{StetsonL} & Variability index describing the synchronous variability of different bands that utilises both \textit{StetsonJ} and \textit{StetsonK}.\\
        \hline
    \end{tabular}
\end{table*}

\begin{table*}
    \caption{Additional light curve derived features implemented in this work. \label{tab:additional_features}}
    \centering
    \begin{tabular}{p{0.25\textwidth}p{0.70\textwidth}}
        \hline
        Feature & Description\\
        \hline
        \textit{median} & Median of magnitudes.\\
        \textit{min\_mag} & Minimum magnitude (maximum brightness).\\
        \textit{max\_mag} & Maximum magnitude (minimum brightness).\\
        \textit{n\_obs} & Number of light curve data points.\\
        \textit{dif\_min\_mean} & Difference between minimum and mean magnitude.\\
        \textit{dif\_min\_median} & Difference between minimum and medium magnitude.\\
        \textit{dif\_max\_mean} & Difference between maximum and mean magnitude.\\
        \textit{dif\_max\_median} & Difference between maximum and median magnitude.\\
        \textit{dif\_max\_min} & Absolute difference between maximum and minimum magnitude.\\
        \textit{temporal\_baseline} & Duration of the light curve.\\
        \textit{pwr\_max} & Maximum power of Lomb Scargle periodogram.\\
        \textit{pwr\_maxovermean} & Maximum over the mean power of the Lomb Scargle periodogram of the light curve.\\
        \textit{npeaks\_\textbf{X}to\textbf{Y}} & Number of peaks with amplitude between \textbf{X} and \textbf{Y}. $\textbf{X}\in (0.5, 1, 2)$ and $\textbf{Y}\in (1,2,5)$. \textit{npeaks\_above5} for peaks above 5 magnitudes.\\
        \textit{rrate\_\textbf{X}to\textbf{Y}} & Maximum rise rate of peaks with amplitude between \textbf{X} and \textbf{Y}.\\
        \textit{drate\_\textbf{X}to\textbf{Y}} & Maximum decline rate of peaks with amplitude between \textbf{X} and \textbf{Y}.\\
        \textit{amp\_\textbf{X}to\textbf{Y}} & Maximum amplitude of peaks with amplitude between \textbf{X} and \textbf{Y}.\\
        \textit{rollstd\_ratio\_t\textbf{A}s\textbf{B}} & Calculate the rolling standard deviation of the light curve with a window size $\textbf{B}\in (5,10)$, where the threshold for the minimum light curve data points, $\textbf{A}\in (10,20)$, is met. The ratio of the highest to lowest standard deviation of these windows is the output.\\
        \textit{stdstilllev\_t\textbf{A}s\textbf{B}} & Ratio of the mean magnitude of the window with the lowest standard deviation to the magnitude range of the light curve - i.e., standstill location relative to the maximum brightness.\\
        \textit{pnts\_leq\_rollMedWin20-\textbf{C}mag} & Number of data points within a rolling window of 20 observations that are brighter than \textbf{C} magnitudes of the median magnitude of that window, where $\textbf{C}\in (1, 2, 5,)$.\\
        \textit{pnts\_geq\_rollMedWin20-\textbf{D}mag} & Number of data points within a rolling window of 20 observations that are fainter than \textbf{C} magnitudes of the median magnitude of that window, where $\textbf{D}\in (1, 2, 3)$.\\
        \textit{pnts\_leq\_median-\textbf{E}mag} & Number of data points brighter than \textbf{E} magnitudes of the median magnitude of the light curve, where $\textbf{E}\in (1, 2, 5)$.\\
        \textit{pnts\_geq\_median-\textbf{F}mag} & Number of data points fainter than \textbf{F} magnitudes of the median magnitude of the light curve, where $\textbf{F}\in (1, 2, 3)$.\\
        \textit{clr\_mean} & Mean of the colours derived at each epoch (night) where an observation in both the g and r band was obtained. Where no epochal colour information is available for a source, the difference between the mean g magnitude and mean r magnitude is used.\\
        \textit{clr\_median} & Same process as used to calculate \textit{clr\_mean}, this time with the median instead of mean magnitude.\\
        \textit{clr\_std} & Standard deviation of the epochal colour.\\
        \textit{clr\_bright} & Colour obtained from epoch where the system is at its brightest. Where epochal colour is unavailable, this is the difference between the minimum g and r band magnitudes.\\
        \textit{clr\_faint} & Colour obtained from epoch where the system is at its faintest. Where epochal colour is unavailable, this is the difference between the maximum g and r band maximum.\\
        \hline
    \end{tabular}
\end{table*}

\subsubsection{Features derived from Gaia} \label{sec:metadata}

In addition to the light curve based features, we also included data from Gaia DR3 \citep{RN545}. From Gaia we made use of G band, red photometer (RP), and blue photometer (BP) filter photometry, including colours and astrometric data (such as parallax and proper motion). Distances and absolute magnitudes are also derived. These supplementary data are included as features that are described in Table \ref{tab:DR3_features}. Such metadata are not available for every source, and we would not expect this information to be available for new sources of unknown class that we wish to classify. We discuss this issue in subsection \ref{sec:missing_data}. 


\begin{table*}
 \caption{Supplementary data from Gaia EDR3 incorporated as dataset features}.
 \begin{tabular}{p{0.40\textwidth}p{0.55\textwidth}}
      \hline
      Feature & Description\\
      \hline
      \textit{ra, dec, ra\_error, dec\_error} & Right ascension, declination, and associated standard errors\\
      \textit{l, b} & Galactic longitude and Galactic latitiude\\
      \textit{ecl\_lon, 'ecl\_lat} & Ecliptic longitude and Ecliptic latitude\\
      \textit{bp\_rp, bp\_g, g\_rp} & BP-RP, BP-G, and G-RP colours\\
      \textit{phot\_\textbf{X}\_mean\_flux} & Mean flux in the G, integrated BP, or integrated RP bands - corresponding to \textit{\textbf{X} = g, bp, or rp} respectively\\
      \textit{phot\_\textbf{X}\_mean\_flux\_error} & Error on the mean flux in the \textbf{X} band\\
      \textit{phot\_\textbf{X}\_mean\_mag} & Mean magnitude in the G, integrated BP, or integrated RP bands - corresponding to \textit{\textbf{X} = g, bp, or rp} respectively\\
      \textit{parallax, parallax\_error} & Gaia parallax in milliarcseconds (mas) and standard error\\
      \textit{pm} & Proper motion (mas/year)\\
      \textit{pmra\_error, pmdec\_error} & Standard error of the proper motion in right ascension and declination directions (mas/year)\\
      \textit{phot\_g\_n\_obs, phot\_bp\_n\_obs, phot\_rp\_n\_obs} & Number of observations in the Gaia G, BP, and RP bands.\\
      \textit{phot\_g\_mean\_mag, phot\_bp\_mean\_mag, phot\_rp\_mean\_mag} & Mean magnitude in the Gaia G, integrated BP and RP bands\\
      \textit{distance} & Distance to the source derived from the inverse parallax (parsecs)\\
      \textit{absmag\_g, absmag\_BP, absmag\_RP} & Absolute Gaia G, integrated BP and RP magnitudes derived from parallax.\\
      \textit{nu\_eff\_used\_in\_astrometry} & Effective wavenumber of the source. Calculated as the photon-weighted inverse wavelength, calculated from the BP and RP spectra ($\lambda^-1$).\\
      \hline
 \end{tabular}
 \label{tab:DR3_features}
\end{table*}

\subsection{Training, validation and test sets}\label{sec: train_test_split}

Supervised classification algorithms require a training dataset for learning patterns and relationships present within the data to generate a model capable of inference. Training set examples are selected from the original dataset, the remainder of which is used for testing of the resultant model. Should the dataset be sufficiently large, a validation set, usually the same size as the test set, will also be obtained. The validation set is used to tune algorithm specific parameters (or hyperparameters) that control how a model is trained, while the test set is held back, taking no part in the training and model tuning process. The size of our dataset is insufficient for a separate validation set, with minority class examples numbering only a few dozen. We therefore opt for a technique designed for such cases, stratified k-fold cross-validation. This involves splitting the training set into k separate subsets (or folds) in a stratified manner - each fold contains the same class proportions as the overall training set. A model is trained on k-1 folds and evaluated, based on a given metric, on the remaining fold (validation fold); this step is repeated until each fold has partaken in the validation process. The metric scores for each of the k models are mean averaged to produce a cross validation score. This technique allows us to maintain an adequately sized training set and serves to assess the consistency of our model (and data). We use a stratified train-test set split ratio of 70:30 and use a 10-fold stratified cross-validation procedure for hyperparameter tuning and model evaluation. The 70:30 split holds back for testing at least a dozen examples for minority classes whilst providing a high proportion of examples for the algorithm to learn patterns during training and for validation.

\subsection{Feature selection} \label{sec:feature_selection}

Our dataset consists of over 250 features, and with only 1439 examples, we introduce the `curse of dimensionality' \citep{RN534}, which refers to a set of problems arising from high dimensionality datasets. As you add dimensions (features) you rapidly increase the minimum amount of samples required to adequately represent all combinations of feature values in your dataset. Increasing the dimensionality increases the complexity of the model whilst also causing the model to become increasingly dependent on the training set, thus leading to overfitting. Selecting the features most informative for our task enables ML algorithms to train faster, reduces complexity allowing for easier interpretation, reduces overfitting, and can improve model accuracy for the right subset of features. To identify the optimal feature subset, the Variance Inflation Factor (VIF; \citealt{RN562}), the one-way Analysis Of Variance (ANOVA; \citealt{RN563}), and the mutual information score \citep{RN563} methods were examined from the filter feature selection family that measures the relevance of features by their correlation with the dependent variable. From the wrapper method family, that examines the usefulness of a subset of features by training a given model on them, the forward feature selection method was chosen. These methods were applied to the training set only to avoid data leakage - information about the target being present in the training set that would not be available when the model is used for prediction \citep{RN560,RN561}.

\subsubsection{Forward Feature Selection}

Forward feature selection (FFS) is an iterative method starting with a model with no features. With each iteration we add a feature, the one that produced the greatest increase in a performance metric as measured on a validation set. The process continues until no further performance increase is measured. The set of selected features may differ based upon the choice of machine learning algorithm (Section \ref{sec:ML algorithms}). Different algorithms often work best with distinct subsets of features, and the method can adapt to these individual requirements.

FFS is utilised for all but the Decision Tree based algorithms (Section \ref{DT based algorithms}) as they naturally determine the most important features during the tree-building process.

\subsubsection{Variance Inflation Factor (VIF)}

VIF is a method used to detect multicollinearity - the existence of a linear relationship between two or more explanatory (independent) variables. It measures how much the variance of the estimated regression coefficients are inflated as compared to when the predictor variables are uncorrelated. It is found by regressing each independent variable on the remaining independent variables to assess the degree to which it is explained by the remaining variables. VIF is given by:

\begin{equation}
    VIF=\frac{1}{1-R^{2}}
\end{equation}

where

\begin{equation}
    R^{2}=1-\frac{SS_{res}}{SS_{tot}}
\end{equation}

where $SS_{res}$ is the sum of squared residuals to the line of best fit in a linear regression model, while $SS_{tot}$ is the sum of squared residuals to the average value. One uses this selection method by iteratively removing features with the highest VIF and recalculating the metric. A VIF equal to 1 represents the absence of multicollinearity, while the effects of multicollinearity increases with increasing VIF. While it is desirable to have VIF as close to 1 as possible, this generally leads to the removal of variables that have a high positive impact on model performance if we are not careful with our implementation of the technique. One must be careful to assure the feature calculation is present in some form within remaining features to maintain the associated information. VIF is particularly beneficial when dealing with feature redundancy that may arise when a feature is derived from both the g and r bands. We experimented with VIF values of 10, 5, 2.5, and 1.5 for all but the Decision Tree based algorithms since decision trees select features in a greedy fashion and make no assumptions on relationships between features.

\subsubsection{One-way ANOVA}

One-way ANOVA compares the mean value of a variable for each of three or more groups. It determines if any of those means are statistically significantly different from each other. The null hypothesis states that there is no statistically significant difference between any two group means: 

\begin{equation}
    H_0 = \mu_1=\mu_2=\mu_3=\mu_4 = ... = \mu_k
\end{equation}

where $\mu$ is a group mean and $k$ is the number of groups. The alternative hypothesis states that at least one of the groups is statistically significantly different from another at a significance threshold of 5\%. This statistic was used to identify the significance of each feature ordered by p-value. A given algorithm was then trained using the top $x$\% of the most significant features and the model cross validation performance recorded. This step was repeated, increasing the values of $x$ in 5\% increments from 5\% to 95\%, to arrive at a subset of features where model performance was strongest. This method is akin to forward feature selection, though with features added based on a statistical test rather than overall model performance. The motivation for the usage of one-way ANOVA lies in its goal to select a set of features that hold significant importance in differentiating between classes. As with FFS and VIF, this approach was only performed with non Decision Tree based algorithms.

\subsubsection{Mutual information}

Mutual information (MI) is the application of information gain (typically used in the construction of decision trees) to feature selection. The MI score measures the degree to which two variables are related. A score of zero is produced if the two variables are independent, and higher values for higher dependencies. For two jointly discrete random variables $x$ and $y$, MI takes the form:

\begin{equation}
    Mutual\;Information = \sum_{x\in X}\sum_{y\in Y} p(x,y) \ln \left[\frac{p(x,y)}{p(x)p(y)}\right]
\end{equation}

We make use of the \textit{scikit-learn} implementation, which uses a nearest neighbour method instead of binning to handle cases where the independent variable (feature), $x$, is continuous, assuming a discrete target, $y$, (see \cite{RN535}). Under the MI feature selection protocol, the most performant features were identified in the same way as for one-way ANOVA, resulting in slight variations in the optimal subset of features for each algorithm. In a similar fashion to one-way ANOVA, MI aims to select features most crucial for class distinction. However, MI quantifies the information shared between features and the outcome, thereby unveiling non-linear, intricate relationships. This feature reduction method was not employed for the Decision Tree based algorithms for the same reasons as above, and furthermore, MI is at the heart of the operation of these algorithms.

\subsection{Class balancing}

A difference in class frequencies affects the predictability of a model. Differences in class prevalence cause algorithms to be biased towards learning patterns more specific to the majority class, and produce models that perform poorly in minority class predictions. To handle the class imbalance present within the dataset (see Table \ref{tab:class numbers}) we tested both a non-sampling method, class weighting (should the algorithm permit such a strategy), and undersampling of the majority class combined with the minority class over-sampling technique Adaptive Synthetic (ADASYN; \citealt{RN537}), a variation of Synthetic Minority Over-sampling Technique (SMOTE \citealt{RN439}).

\subsubsection{Class weighting}

Rather than augmenting the dataset, one may modify the algorithm to account for skewed class distributions by giving different weights to each class depending on their dataset prevalence. The difference in weights influences the classification during the training phase. The goal is to penalise the miss-classification of the minority class by setting a higher class weight, while at the same time reducing the majority class weight. Weightings are applied within the cost function for each algorithm such that the miss-classification of a minority class example (e.g., an AM CVn) leads to a greater cost penalty than for a majority class example (e.g., an SU UMa).

\subsubsection{ADASYN}

SMOTE works by selecting a random example from the k nearest neighbours in feature space of a randomly chosen example from the minority class (or class of choice); draws a line in this feature space between the examples and generates a new sample at a random point along that line. The ADASYN adaptation generates more synthetic examples in regions of feature space where the density of minority examples is low, and fewer or none where the density is high. Subsequently, more synthetic data is generated for minority class samples that are harder to learn compared to those where many examples are available, making it easier to learn.

\subsection{Missing Data} \label{sec:missing_data}

Missing data due to insufficient data points during the light curve feature extraction process accounts for as much as 20\% for a given feature. Whilst that due to unavailability of metadata accounts for up to 33\%. Many machine learning algorithms do not support missing values, therefore strategies must be implemented to address this absence of data. The most common and simplest strategy is to replace (or impute) missing values with the mean or median of the feature, however, this method ignores relationships between features and reduces the variance of the variable, thereby introducing bias to the model. The following aims to mitigate such bias.

Adopted here is a two step approach, firstly the reasons for missingness is assessed, and we assign either an appropriate value, such as that for the other filter, a value based on an immediately relevant feature, or the value for that feature is left as missing. The final step is to utilise the scikit-learn implementation of the K Nearest Neighbour imputation method \citep{RN542}. For each sample, each missing feature is imputed using the values from the k nearest (based upon some distance metric, typically euclidean) neighbours in feature space where that feature value is present. The imputed value will be either the uniform or weighted-by-distance average feature value for those neighbours. We implement this method using the weighted-by-euclidean distance average for imputation with the default five nearest neighbours.

\subsection{Machine Learning algorithms}\label{sec:ML algorithms}

The algorithms whose performance we evaluate are \textit{scikit-learn's} \citep{RN457} Python implementations of Random Forest (RF) \citep{RN279}, K-Nearest neighbours (KNN) \citep{RN447}, Gaussian Naive Bayes \citep{RN538}, and Linear Discriminant Analysis (LDA; \citealt{RN539}). Also used are the Extreme Gradient Boosting (XGBoost) algorithm \citep{RN295} and Keras \citep{chollet2015keras} 
implementation of an Artificial Neural Network (ANN) in the form of a Multi-Layer Perceptron - a fully connected multi-layer ANN \citep{RN483}. Furthermore, for model evaluation and interpretability purposes we used \cite{RN513} python implementation of Generative Topographic Mapping \citep{RN512}.

The array of algorithms embody a diverse spectrum of classification strategies chosen to extract optimal insights from the dataset. RF is adept at navigating intricate patterns in data through its ability to handle non-linear relationships, high-dimensional data, and noisy features. XGBoost is known for delivering high-performance scalability, often surpassing other algorithms and underscoring the potential of ensemble methods. KNN adds instance-based learning to the mix, Naive Bayes adds probabilistic modeling, and LDA is adept at discerning linear separability. Meanwhile, the multi-layer perceptron, is a fundamental deep learning architecture, these are capable of capturing intricate patterns in data.

\subsubsection{Decision Tree based ensemble methods} \label{DT based algorithms}

RF and XGBoost are built with an ensemble of Decision Trees (DT; \citealt{RN278}) combined either non-sequentially or sequentially, respectively. With the provided features, DTs employ a series of binary splits on the training dataset, starting from the root node. These splits aim to create groups (referred to as leaf nodes) that maximise dissimilarity while moving closer to a uniformity of class within each group. The resulting model utilises this hierarchical tree structure to make predictions on unseen instances. 

RF operates by employing a voting mechanism, using predictions generated by a randomly selected set of DTs. The class with the highest number of votes becomes the prediction of our model. Using the bootstrap aggregation technique, each tree is trained on a modified variant of the original training set. Additionally, a random subset of features is used during this process to ensure the trees remain uncorrelated. Several crucial hyperparameters come into play. Increasing the number of trees enhances the model's ability to generalise, albeit at the expense of added complexity and computational time. Adjusting the maximum tree depth, which dictates the furthest distance from the root to a leaf node, and altering the maximum number of features available to each tree serve to control against overfitting.

XGBoost employs a boosting approach that makes predictions for \textit{n} rounds on the training sample, iteratively improving its performance with each round by utilising information from the prior round’s prediction accuracy. Specifically, its goal is to minimise a loss function by iteratively selecting a tree that points towards the negative gradient of the said function. XGBoost utilises parallelised tree building and hardware optimisation to improve runtime, and regularisation to reduce overfitting. Hyperparameters include those mentioned for RF with the addition of parameters such as the learning rate that controls the loss function step size at each iteration, and the regularisation rate to adjust model generalisation.

\subsubsection{K Nearest Neighbours}

KNN \citep{RN447} stores the feature-space position vectors of training set examples. When making class predictions for new examples, it identifies the mode of the classes among the k nearest neighbors from the training set, assigning that mode as the prediction for the new example. The hyperparameters that impart the greatest influence on model performance are the number of nearest neighbours, the distance metric for similarity computation, and the weighting of individual examples. The algorithm was implemented and evaluated using the complete set of features, as well as with subsets of features determined by the feature selection methods detailed in Section \ref{sec:feature_selection}.

\subsubsection{Artificial Neural Networks}

Artificial Neural Networks (ANN) \citep{RN448}, comprise interconnected layers of nodes, commonly referred to as neurons. This architecture consists of an input layer that receives feature values, an output layer responsible for generating predictions, such as class probabilities, and one or more hidden layers in between. The hidden layers sequentially transform the initial feature values into predictions by applying non-linear functions to linear combinations of previous inputs. The learning process revolves around minimising a loss function, where adjustments to the model parameters are made through an iterative process known as backpropagation until convergence to loss minimum is achieved. ANN implementation and evaluation follows the same feature selection methodology as for KNN, albeit excluding FFS due to its impracticality and computational expense when applied to ANN.

\subsubsection{Linear Discriminant Analysis}

Linear Discriminant Analysis (LDA) is a dimensionality reduction technique also used for classification purposes. Class predictions are obtained by finding the class that maximises the posterior probability from Bayes' rule. Class distributions are modelled as multi-variate Gaussians assumed to have the same covariance for each class. This assumption reduces the log of the posterior probabilities to linear functions, which leads to a further assumption, linear separability, since locations where the functions are equal define linear class decision boundaries. LDA is evaluated with both the full feature set and the subsets of features determined by the methods of Section \ref{sec:feature_selection}.

\subsubsection{Gaussian Naive Bayes}

As with LDA, the Gaussian Naive Bayes (GNB) classifier is based on Bayes' theorem, though unlike LDA, a naive assumption is made; the features are independent, that is the presence of one feature has no effect on others. The class y that gives the maximum posterior probability is assigned to a given example. The Gaussian arises because the predictors take on continuous values, and are considered to be sampled from a Gaussian distribution. The features utilised in GNB follow the same methodology used for feature selection as applied to KNN and LDA.

\subsubsection{Generative Topographic Mapping} \label{sec:gtm}

GTM is a neural network based manifold learning algorithm that is able to compute a mapping between points in low dimensional (often 2D) latent space into a higher dimensional data space \citep{RN512}. This is performed such that the structure of the latter is represented in the former, in other words, points close together in latent space will map to points close together in data space. Points in latent space are arranged in an equally spaced grid of nodes. GTM performs a non-linear mapping between those points and points in data space using a linear combination of radial basis functions with weighting coefficients. These points in data space represent the centres of Gaussian probability density functions that make up a mixture of Gaussians. During training, the weights and variances are adjusted using the Expectation Maximisation algorithm such that the overall probability distribution of data space is accurately represented by the Gaussian mixture. The Gaussian centres will converge to the mean or median locations of local structures (clusters) in data space. It is with this overall probability density function defined by the mixture that examples can be mapped to locations in latent space based on the responsibilities of Gaussians in the mixture.

One may utilise GTM to evaluate the ability of a classifier to distinguish between classes, and to identify the features responsible for the assignment of a given class rather than an overall feature importance list that only provides the features responsible for overall model performance. To do this we input the posterior class probabilities for the training set output by our classifier into the GTM framework. Therefore the data space is a class probability space of nine dimensions. Each example from the training set will have posterior probabilities of belonging to each class evaluated by the classifier, these probabilities define their location in class probability space. Distinct clusters of these examples located in regions with high probability along a particular probability space dimension would represent a classifier that can accurately distinguish between classes. Since these clusters define the Gaussian centres, they are mapped to the corresponding nodes in latent space. We can then evaluate this class separability within the latent space representation. This representation forms a grid of squares, each defining a node, colour-coded based on the location of the associated probability space Gaussian centre along a given probability space axis (or particular class probability). 

For feature responsibilities we simply average a particular feature value for all examples assigned to a given node, assigned meaning the node with the highest likelihood of being responsible for a given example. The average for each node can then be used to produce a 2D histogram consisting of the same above latent space grid with squares colour coded by these averages, one for each feature. The distribution of mean feature values can be analysed against the distribution of classes in the class maps to identify class specific features.

\subsection{Performance Metrics} \label{sec:performance_metrics}

Performance metrics rely on the counts of true positives (TP), true negatives (TN), false positives (FP), and false negatives (FN). To compute these counts, one must establish the positive class, representing the class of interest (e.g., one of the CV classes), and the negative class, encompassing all other classes in this multi-class scenario.

Frequently, the counts of TP, TN, FP, and FN are organised in an N $\times$ N table referred to as a \textit{confusion matrix}, with N signifying the number of classes. This matrix provides a straightforward means to view the quantities of TPs, TNs, FPs, and FNs. These values are used to calculate the class-specific precision, recall, and F1-score, as well as the balanced accuracy and Area under the Curve of the Receiver Operating Characteristic.

The precision defines the proportion of instances our model predicts as belonging to the positive class, and actually, do belong to this class: $TP/(TP+FP)$. It offers insights into the reliability of our model's predictions. Recall, on the other hand, measures the fraction of positive class instances correctly predicted by our model: $TP/(TP+FN)$. This metric assesses the model's capacity to identify all members of the positive class. The F1-score represents the harmonic mean of precision and recall for the class of interest. It is useful in finding the best trade-off between these quantities. A perfect F1-score is 1 (100\%), indicating both perfect precision and recall, while the lowest possible value is 0, suggesting a score of 0 for either precision or recall. Balanced accuracy, unlike the basic accuracy metric (which is the ratio of the sum of true positives for each class to the total number of examples), calculates the arithmetic mean of recalls for each class. This is particularly useful when dealing with class imbalances, where the basic accuracy metric may not accurately reflect model performance.

The Receiver Operating Characteristic (ROC) curve offers a visual representation of the trade-off between sample purity and completeness. It plots the True Positive Rate (TPR), also known as recall, against the False Positive Rate (FPR). The FPR represents the fraction of examples incorrectly classified as belonging to the positive class, calculated as $FP/(TN+FP)$. This curve is generated by varying the threshold probability used to determine positive classifications for each example. In detail, ML algorithms provide a class probability score for each example, and a threshold is applied to classify examples as positive or negative. The ROC curve showcases the performance of the TPR and FPR as this probability threshold is continuously adjusted. This tool is valuable for selecting an appropriate threshold that aligns with the desired balance between purity and completeness, depending on the specific research objectives. In classification tasks, the goal is to maximise TPR while minimising FPR. An area under the curve (AUC) value of 1 indicates a perfect model that correctly assigns class predictions for all examples. An AUC of 0.5 signifies a model no better than random guessing, while an AUC of 0 implies incorrect predictions for all examples. Although ROC curves are typically associated with binary classification, in this case of multi-class models, they are generated using a one-versus-rest approach. This entails designating one class as the positive class and the remaining classes as the negative class to produce separate curves for each class.

While such performance metrics can be used to assess test set performance differences between classifiers, the McNemar's test can be utilised to judge statistically significant differences between the test set predictions of any two classifiers.  The null hypothesis states that the classifiers disagree in their class predictions to the same amount. Should this be rejected, the alternative hypothesis implies there is evidence they disagree in different ways. The test statistic is calculated in the following way:

\begin{equation}
    statistic=\frac{(Yes/No-No/Yes)^{2}}{Yes/No + No/Yes}
\end{equation}

where $Yes/No$ is the number of test instances that classifier 1 got correct and classifier 2 got incorrect, while $No/Yes$ describes the opposite of this. The test statistic follows a chi-squared distribution with one degree of freedom. The test is usually administered in a binary classification setting, however, under the multi-class case, the correct and incorrect classifications are performed for each class.

\section{Results} \label{sec:results}

\subsection{Classifiers}

\begin{figure*}
    \centering
    \includegraphics[width=\textwidth]{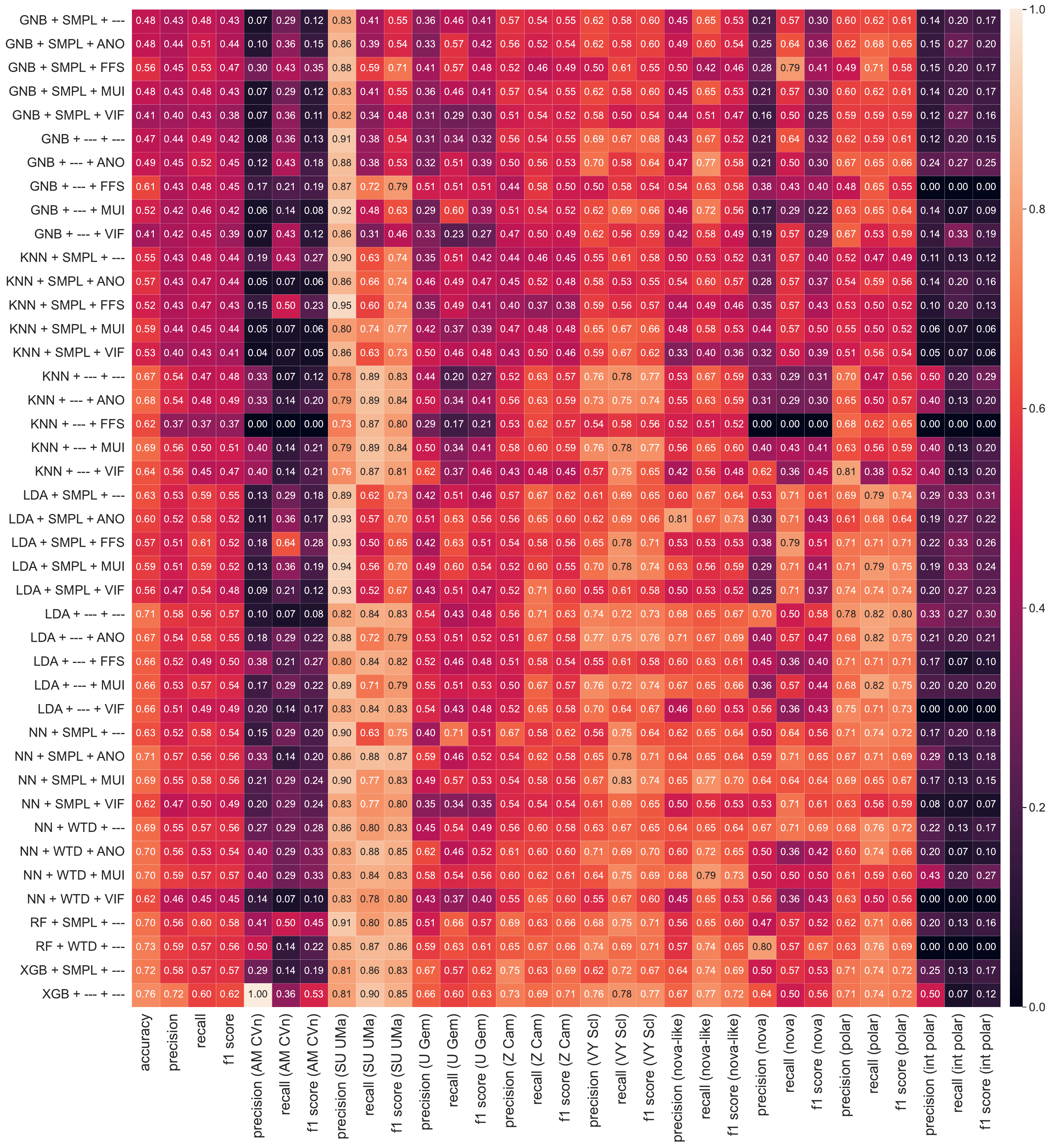}
    \caption{Presented as a heatmap are, the accuracy, and the macro average quantities of precision, recall, and F1-score for each classifier variant. Alongside these are the precision, recall, and F1-score for each class. Classifiers are labelled as follows: classifier + class balancing method + feature selection method. Classifier abbreviations are as described in the text, the class balancing methods are abbreviated as SMPL, WTD, or ---, depending on whether over/under sampling methods, class weighting, or no class balancing method was implemented, respectively. Feature selection methods are abbreviated as ANO, FFS, MUI, VIF, or ---, for one-way ANOVA, forward feature selection, mutual information, variance inflation factor, or no such implementation (full set of features used), respectively.}
    \label{fig:results_table}
\end{figure*}

In our study, to distinguish between the nine CV classes, we evaluated several algorithms: Gaussian Naive Bayes, Linear Discriminant Analysis, K-Nearest Neighbors, Random Forest, XGBoost, and a multi-layer perceptron neural network. To address class imbalance, we used either the class weighting method (where possible) or the ADASYN oversampling technique in combination with random undersampling to balance the training set. Training was conducted on subsets of features determined through the mutual information score, variance inflation factor, the one-way ANOVA method, or forward feature selection. We assessed the resultant models based on overall accuracy, macro averages of precision, recall (equivalent to balanced accuracy for the macro average), and F1-score. These are provided in the heatmap shown in Figure \ref{fig:results_table} within the first four columns. The corresponding precision, recall, and F1-scores for each class are provided in the remaining columns.

To compare the test set performance metric means of different classifier groups, we conducted T-tests. The results indicate that GNB and KNN-based classifiers performed poorly on the test set compared to the other algorithms (F1-score of $0.44\pm0.04$ and $0.54\pm0.04$, respectively, with a p-value of $p=8.6\times10^{-11}$ at $\alpha=0.05$). However, there was no significant performance difference when using over/under sampling compared to class weighting (or no such method) ($p=0.07-0.86$ for all metrics). Regarding feature reduction methods, we observed small but not significant performance improvements when using the one-way ANOVA and mutual information, while the use of variance inflation factor led to a performance drop.

The class-specific performance associated with each model revealed difficulties in correctly classifying the AM CVn and intermediate polar classes, irrespective of the algorithm used or any effort to address class imbalance. These two classes, along with the nova class, have the lowest sample size. Despite the small sample size, the light curves (and metadata) of the novae are sufficiently distinct for the algorithms (especially NN, RF, and XGBoost) to distinguish them from the remaining classes. All models, except for GNB, performed well in classifying the SU UMa class. 

\begin{figure*}
    \centering
    \includegraphics[width=\textwidth]{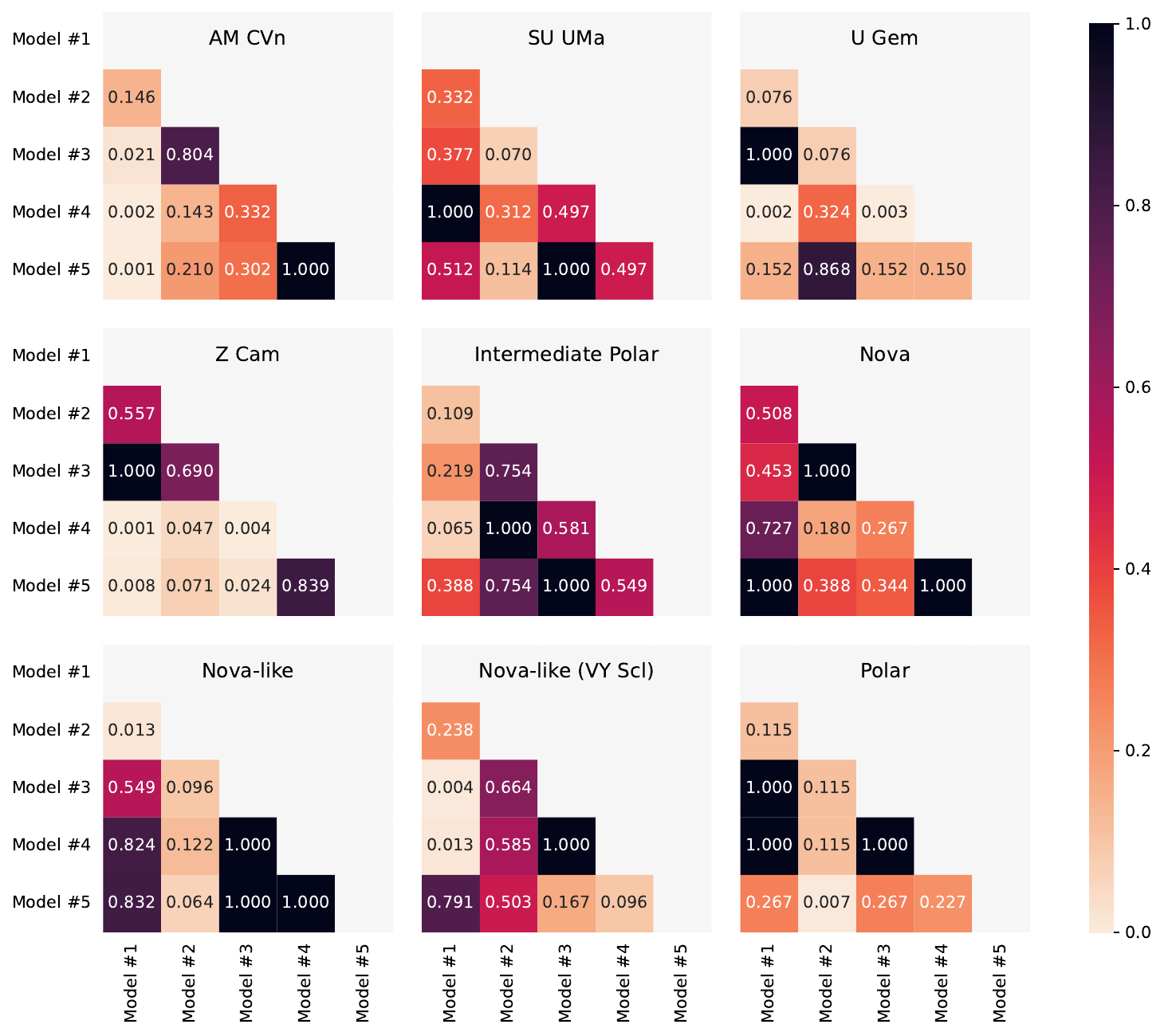}
    \caption{
    The per-class p-values from McNemar's tests conducted between each pair of the top 5 ranked classifiers from Table \ref{tab:top5}. For ease of reference these are, from rank 1 to 5, XGB + --- + ---, RF + SMPL + ---, XGB + SMPL + ---, NN + WTD + MUI, AND LDA + --- + ---. The significance threshold is set to p=0.05, the classifier descriptions and abbreviations are as described in the caption of Figure \ref{fig:results_table}.}
    \label{fig:McNemars}
\end{figure*}

To select the model for the pipeline, we based our decision on the macro F1-score with some consideration for the performance on the lowest sample size classes. Table \ref{tab:top5} presents the top 5 models based on the macro F1-score, while Figure \ref{fig:McNemars} shows the class-specific 'p-value table' resulting from a McNemar's test for each pair of these models. The figure indicates no significant prediction disagreements between these algorithms for the SU UMa, nova, and intermediate polar classes. However, models ranked in the top 3 show significant prediction disagreements compared to models ranked 4 and 5 with regards to Z Cam. For the AM CVn class, the XGBoost classifier, implemented without explicit class balancing or feature reduction, significantly outperformed the other models. As a result, we selected this XGBoost model trained with 500 decision trees at a maximum tree depth of 14, as the classifier for the second stage of our pipeline.

We should note that certain aspects of the model selection, such as the variation of examples apportioned to the training, validation, and test sets, the NN weights initialisation, and the feature selection for each tree of the RF and XGB models, were randomly selected. Thus, different random initialisations could have led to the selection of any of the models generated from the NN, RF, and XGB algorithms.

\begin{table}
    \centering
    \caption{Top 5 ranked classifiers based on the macro-averaged F1-score. Listed are the algorithm, the method used to handle class imbalance and the method used to reduce the number of features. The class balancing methods are abbreviated as SMPL, WTD, or -, depending on whether over/under sampling methods, class weighting, or no class balancing method was implemented, respectively. The only feature selection methods in this list are those abbreviated as MUI or -, for mutual information or no feature selection method (full list of features used), respectively.}
    \label{tab:top5}
    \begin{tabular}{c|l|c|c|c}
        \hline
        Rank & Algorithm & Imbalance & Feature selection & F1-score\\
        \hline
        1 & XGB & ---  & --- & 0.62 \\
        2 & RF  & SMPL & --- & 0.58 \\
        3 & XGB & SMPL & --- & 0.57 \\
        4 & NN  & WTD  & MUI & 0.57 \\
        5 & LDA & ---  & --- & 0.57 \\
        \hline
    \end{tabular}
\end{table}

\subsection{Performance} \label{sec:performance}

\begin{figure*}
    \centering
    \includegraphics[width=\textwidth]{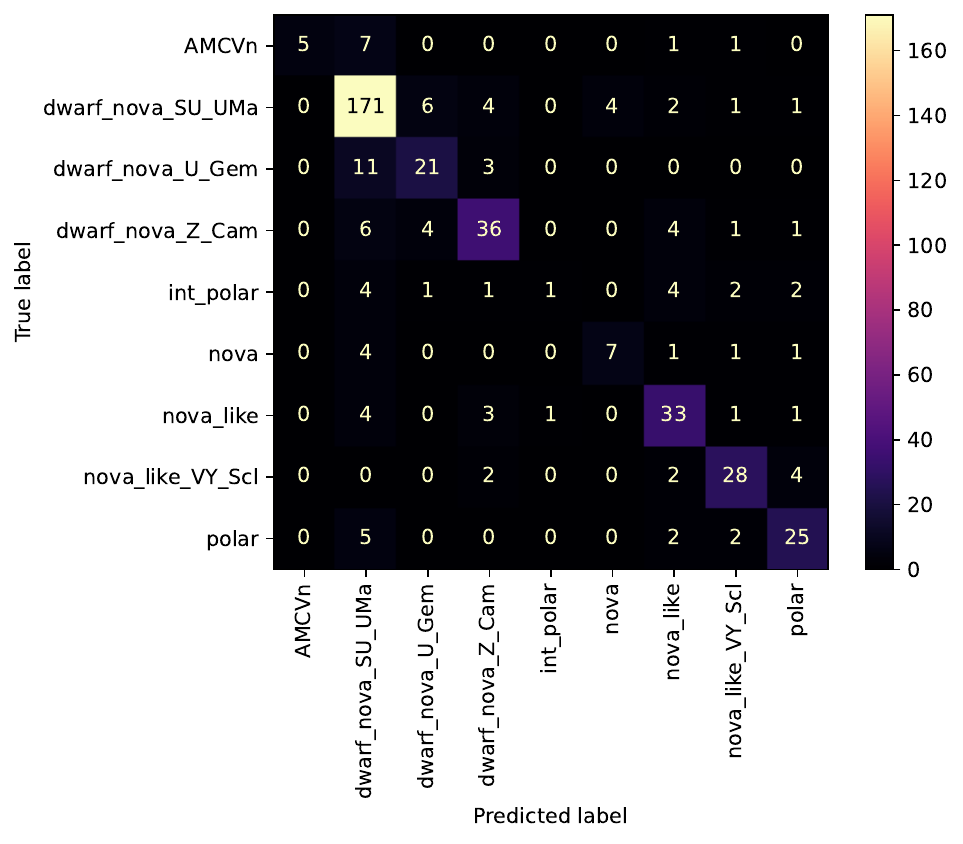}
    \caption{Confusion matrix for the XGBoost model.}
    \label{fig:confusion_matrix_XGB}
\end{figure*}

The per class performance of the model as implemented on the test set is described in Table \ref{tab:classification_report_XGB}, while the corresponding confusion matrix is shown in Figure \ref{fig:confusion_matrix_XGB}. Evident are the following. SU UMa is responsible for the highest precision and recall scores, contributing greatly towards an increase in the overall classification performance, Z Cam and VY Scl are also well picked out by our classifier. The overall performance suffers noticeably due to the performance of the intermediate polar class. Intermediate polars represent a class subject to one of the largest amount of training set oversampling, due to a low number of examples.

\begin{table}
    \caption{Classification report for the XGBoost model. For each class of CV the precision, recall, F1 score, and the number of test set examples are given. The macro average (or arithmetic mean) of each metric, accuracy and balanced accuracy are also provided.\label{tab:classification_report_XGB}}
    \centering
    \begin{tabular}{l|r|r|r|r}
        \hline
        Class & Precision & Recall & F1 score & Test set amount\\
        \hline
        AM CVn & 1.00 & 0.36 & 0.53 & 14\\
        SU UMa & 0.81 & 0.90 & 0.85 & 189\\
        U Gem & 0.66 & 0.60 & 0.63 & 35\\
        Z Cam & 0.73 & 0.69 & 0.71 & 52\\
        Intermediate Polar & 0.50 & 0.07 & 0.12 & 15\\
        Nova & 0.64 & 0.50 & 0.56 & 14\\
        Nova-like & 0.67 & 0.77 & 0.72 & 43\\
        Nova-like VY Scl & 0.76 & 0.78 & 0.77 & 36\\
        Polar & 0.71 & 0.74 & 0.72 & 34\\
        \hline
        Macro average & 0.72 & 0.60 & 0.62 & 432\\
        Accuracy & & & 0.76 & 432\\
        Balanced accuracy & & & 0.60 & 432\\
        \hline
    \end{tabular}    
\end{table}

Also falling within this high oversampling bracket are the AM CVns and novae. Despite this, they are responsible for strong precision scores such that 100\% of examples predicted as AM CVn and 64\% of examples predicted as nova are true members of the class. However, this does comes at the expense of lower recall scores, 0.36 for AM CVns and 0.50 for novae. Those true AM CVn members that are missclassified are mostly assigned the SU UMa class, as are true members of the nova class.

The classifier performs well in distinguishing between systems that regularly display dwarf nova outbursts (where we exclude intermediate polars) from those that do not. Should we group those classes into those that exhibit these outbursts and those that do not, the precision and recall scores for the dwarf nova exhibiting class would be 0.92 and 0.94 respectively, while for non-dwarf nova exhibiting systems, 0.88 and 0.83. Confusion between dwarf nova exhibiting systems is an area where the model performance suffers. Notable is the mislabelling of AM CVn members as SU UMa; and the contamination of predictions of the U Gem class by SU UMa and Z Cam members. Similarly confusion between non-dwarf nova exhibiting system also plays a factor: true intermediate polar members are confused for nova-likes, VY Scl and polars; and confusion between the nova-like, VY Scl and polar classes is present. Reverting our description of performance back to our 9 class problem, notable is the significant missclassification of true Z Cam members with the nova-like class and the significant contribution of false positive by the SU UMa class towards the predictions of the the nova class.

\begin{figure*}
    \centering
    \includegraphics[width=\textwidth]{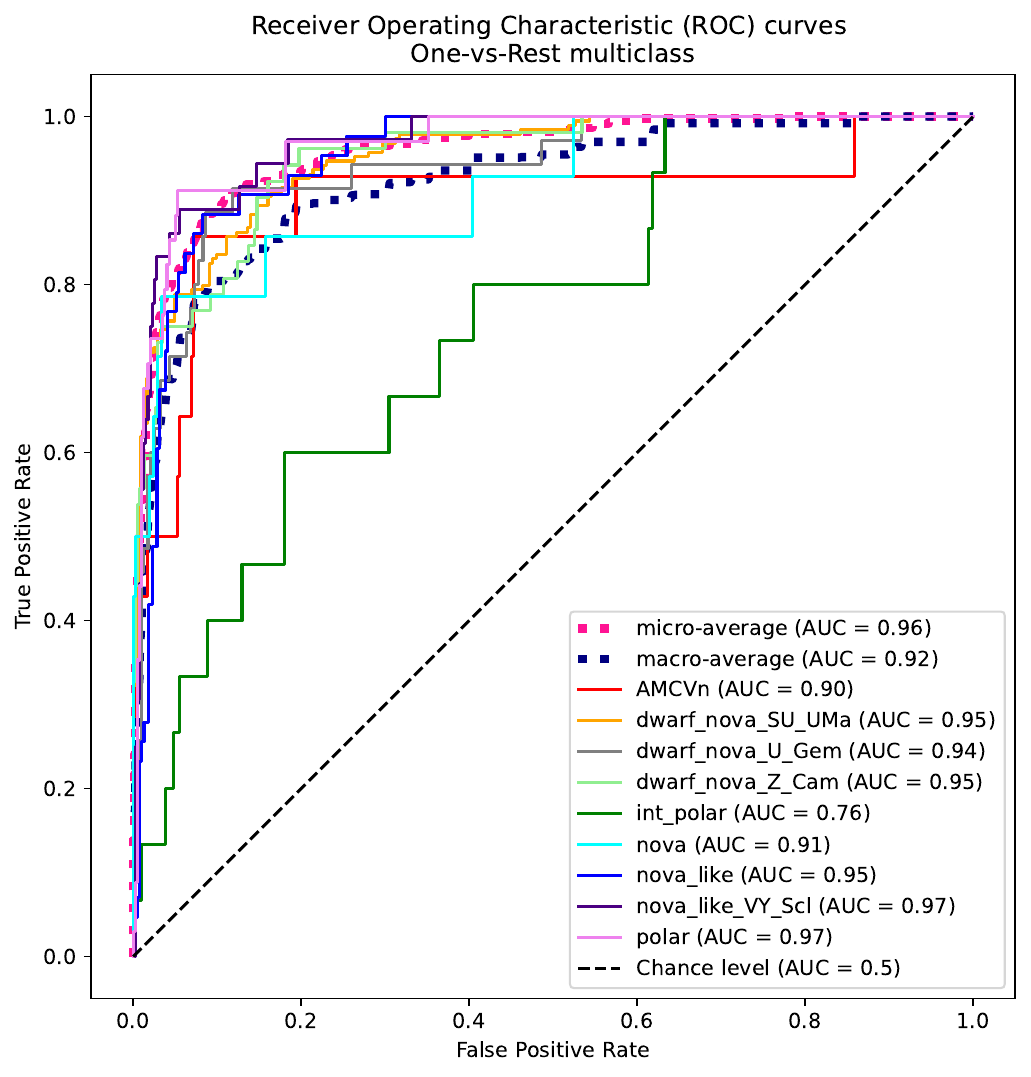}
    \caption{Receiver operating characteristics for the XGBoost classifier.}
    \label{fig:roc_curves}
\end{figure*}

With respect to the ROC Curves Figure \ref{fig:roc_curves}, in all cases the classifier performed much better than a random guess, depicted by the 'chance level' line. An AUC score above 0.93 for all but the intermediate polar and AM CVn classes represents a strongly performing classifier, where the resultant micro and macro averages are 0.96 and 0.92. This is a further illustration of the findings within the confusion matrix and classification report.

\begin{figure*}
    \centering
    \includegraphics[width=\textwidth]{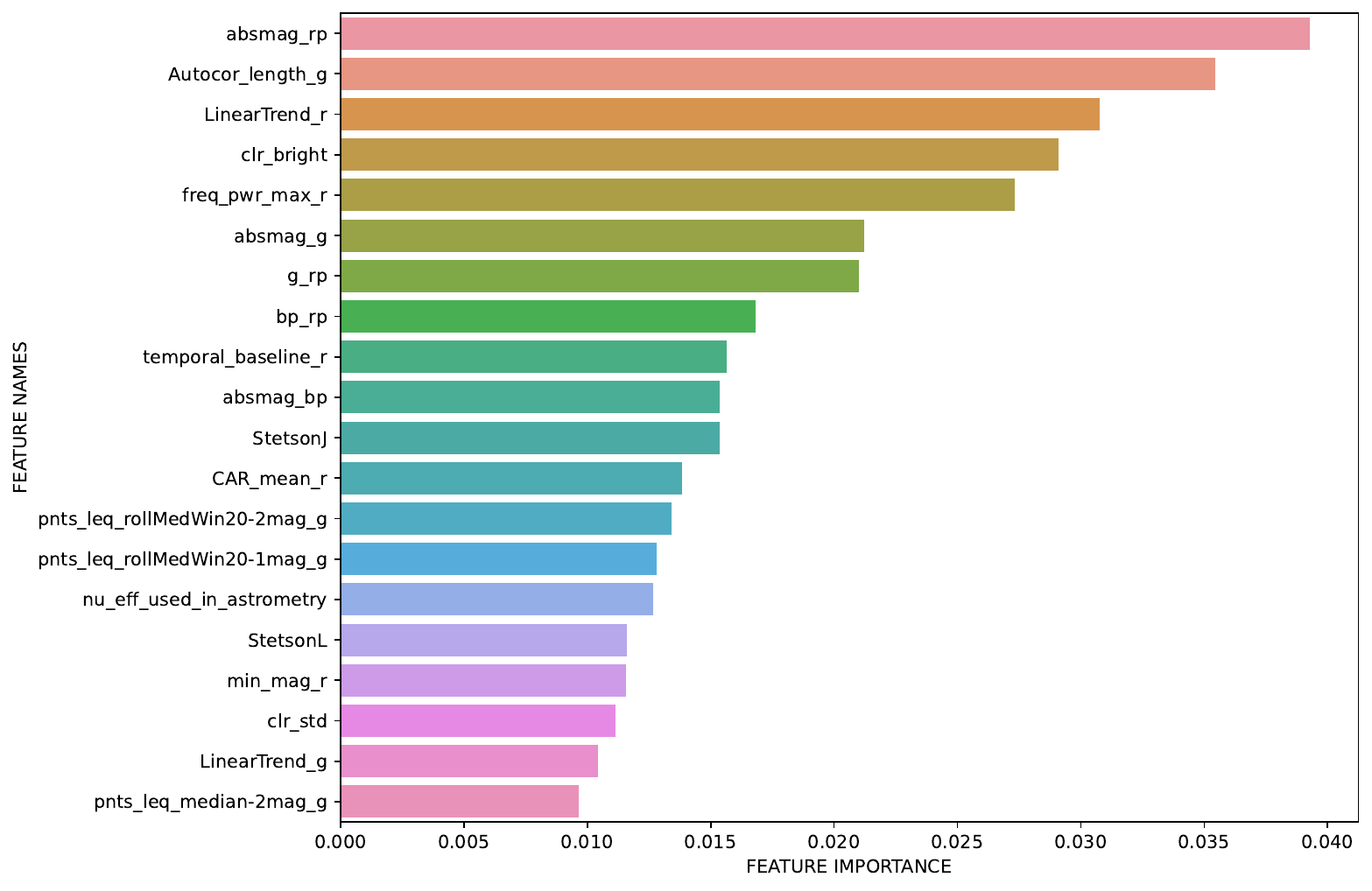}
    \caption{Feature importance scores for the 20 most influential features within the chosen classifier model. Feature importance refers to a class of techniques for assigning scores to input features to a predictive model, indicating the relative importance of each feature when making a prediction.}
    \label{fig:feature_importances_XGB}
\end{figure*}

The importance of each feature for DT-based models can be given by the feature importance scores. The 20 features with the largest effect on the model's predictive accuracy are plotted in Figure \ref{fig:feature_importances_XGB}. Ranked highest is the Gaia RP band absolute magnitude (abs\_mag\_rp); Gaia BP and G absolute magnitudes also feature within the list. ZTF and Gaia colours feature strongly, with the brightest epochal colour (clr\_bright), Gaia G-RP and Gaia BP-RP colours within the top 10. The slope of a linear fit to the ZTF r band light curve is deemed highly relevant for classifier performance, as is the auto-correlation length in the ZTF g band. Periodicity based features within the list come in the form of the frequency of maximum power in the Lomb Scargle periodogram of the r band light curve. Features for identifying outbursts are represented by the number of points brighter than the rolling median. Features that test for the synchronous light curve variability across both bands come in the form of \textit{StetsonJ} and \textit{StetsonL} (see Table \ref{tab:lc_features}). The list therefore contains a mixture of features that cover periodicity, photometry, and statistical descriptors.

\subsection{GTM Latent space representations} \label{sec:gtm_representations}

\begin{figure*}
    \centering
    \includegraphics[width=\textwidth]{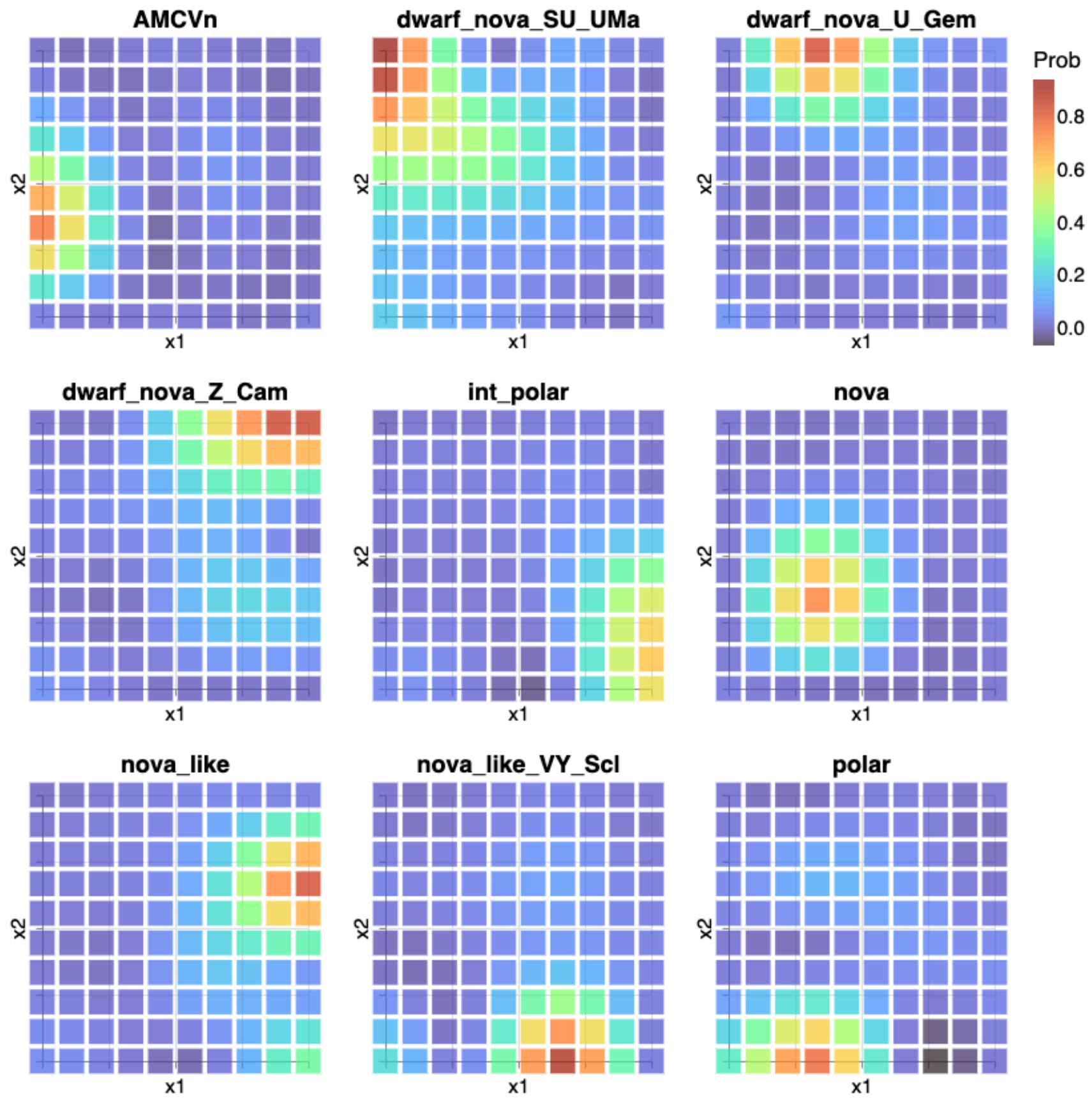}
    \caption{GTM latent space visualisation of the class posterior probability space from the XGBoost classifier chosen for the pipeline.}
    \label{fig:latent_space_visualisation}
\end{figure*}

Class maps generated using GTM, as described in subsection \ref{sec:gtm}, are presented in Figure \ref{fig:latent_space_visualisation}. These latent space representations of class probability space structures assess the class separability of our ML model. The class maps clearly show the existence of structures that are located in fairly distinct regions, each associated with a particular class. This is representative of a classifier that has effectively learnt patterns within the data necessary for class distinction. These structures are extended, with their cores represented by the highest probability of belonging to the associated class, whilst as we move away from the cores, the probabilities diminish (represented by the colour scale). Structures extend into regions associated with that of other classes, indicating some class confusion, thus reflecting observations within the confusion matrix. The highest class probabilities are associated with the SU UMa, U Gem, Z Cam, nova-like, and VY Scl classes - their structure cores exceed 0.80 in class prediction probability. Structures for the AM CVn, nova and polar classes are also present, though with class probabilities no higher than 0.7 and 0.8 respectively. As mirrored in the confusion matrix, the intermediate polar structure, though located in a relatively distinct region, is only responsible for a core class probability of 0.62.

Another interesting feature of the maps are that outbursting systems tend to reside along the top edge and down the left edge, while systems that are not expected to display dwarf nova outbursts are located along the right and bottom edges of the maps. This concurs with our the observation of the effectiveness of our model to distinguish outbursting from non-outbursting systems. The nova class is the only one located away from any edge.

The most obvious blending between structures (or equivalently, confusion between classes) is evident for the SU UMa class - the most prevalent class in the dataset. Its structure extends well into the AM CVn and U Gem regions, also coming into contact with Z Cam and nova. Z Cam is responsible for a well defined structure (top right) that extends into nova-like class probability space, and a tenuous one ($\sim$0.2 in Z Cam class probability) that is more strongly associated with the nova-like, VY Scl and intermediate polar classes. Nova-likes are also responsible for a tenuous, secondary structure (bottom right) more strongly associated with intermediate polars. There is also clear overlap between nova-like, VY Scl and intermediate polar classes, and structure blending in evident between AM CVn, nova, and polar classes.

\begin{figure}
    \centering
    \includegraphics[width=\columnwidth]{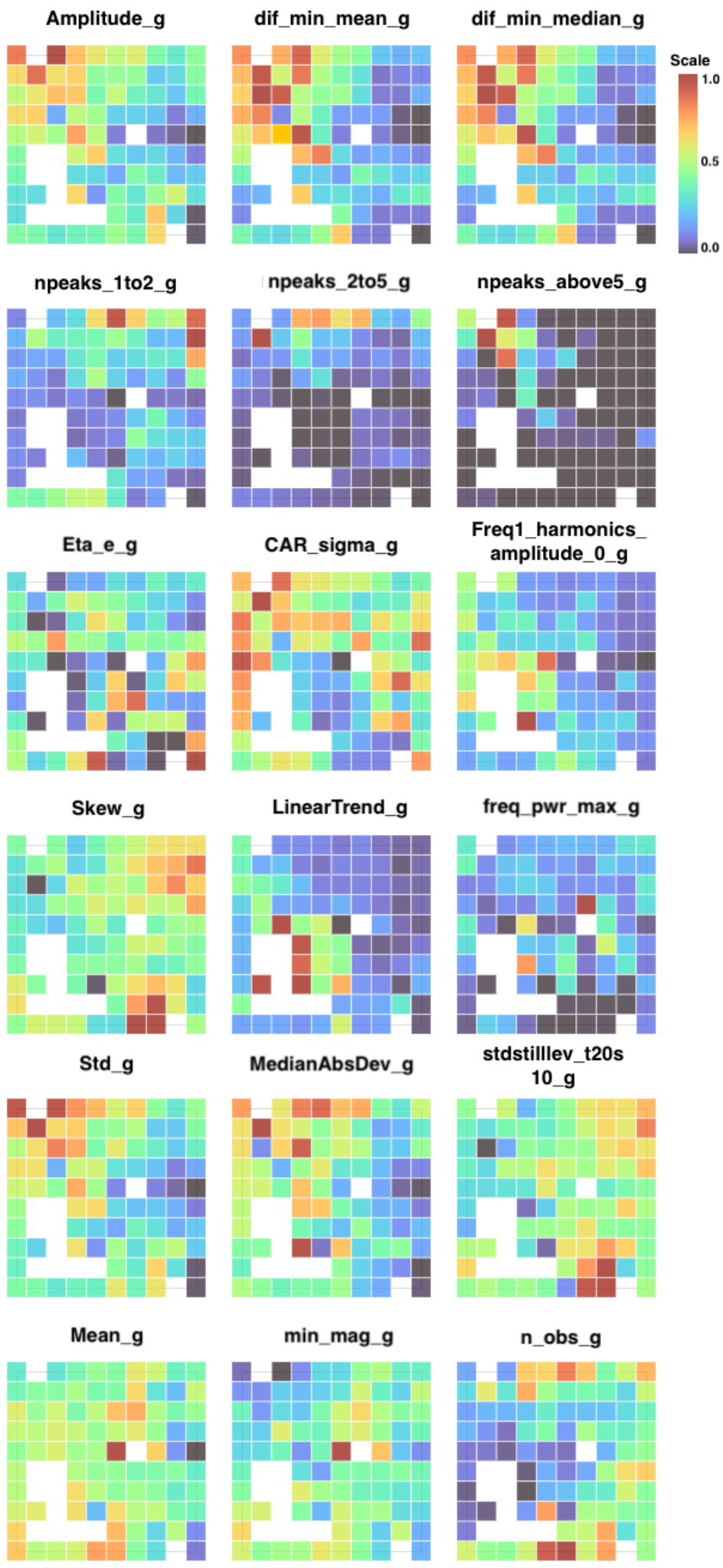}
    \caption{GTM-generated feature maps for the XGBoost model. Compare high and low-value regions to class maps to pinpoint key features for class assignment. White squares indicate empty nodes, to which no examples are assigned, determined by node responsibility (see Section \ref{sec:gtm}).}
    \label{fig:factor_map_g}
\end{figure}

\begin{figure}
    \centering
    \includegraphics[width=\columnwidth]{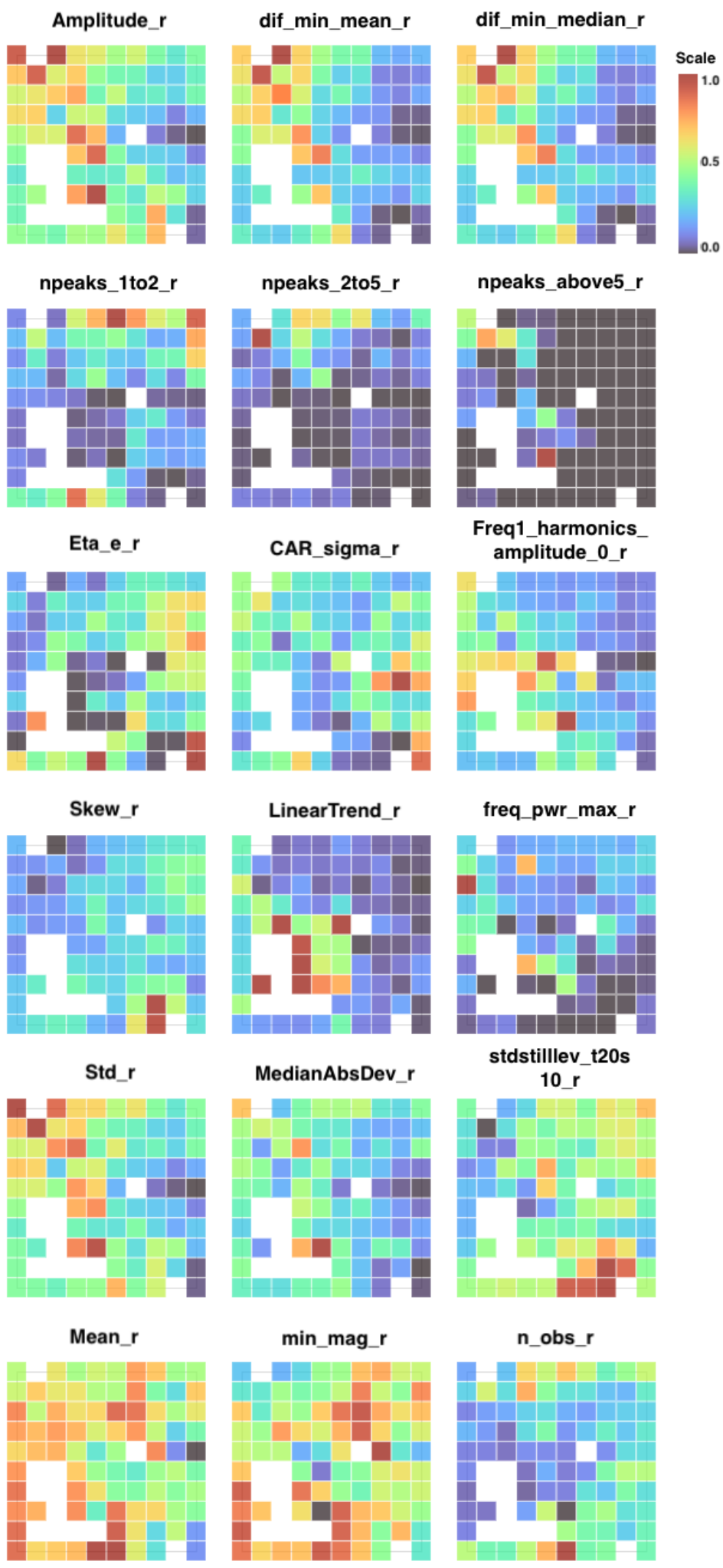}
    \caption{Feature maps for the XGBoost model produced using GTM. Same as for figure \ref{fig:factor_map_g} though for r band}
    \label{fig:factor_map_r}
\end{figure}

\begin{figure}
    \centering
    \includegraphics[width=\columnwidth]{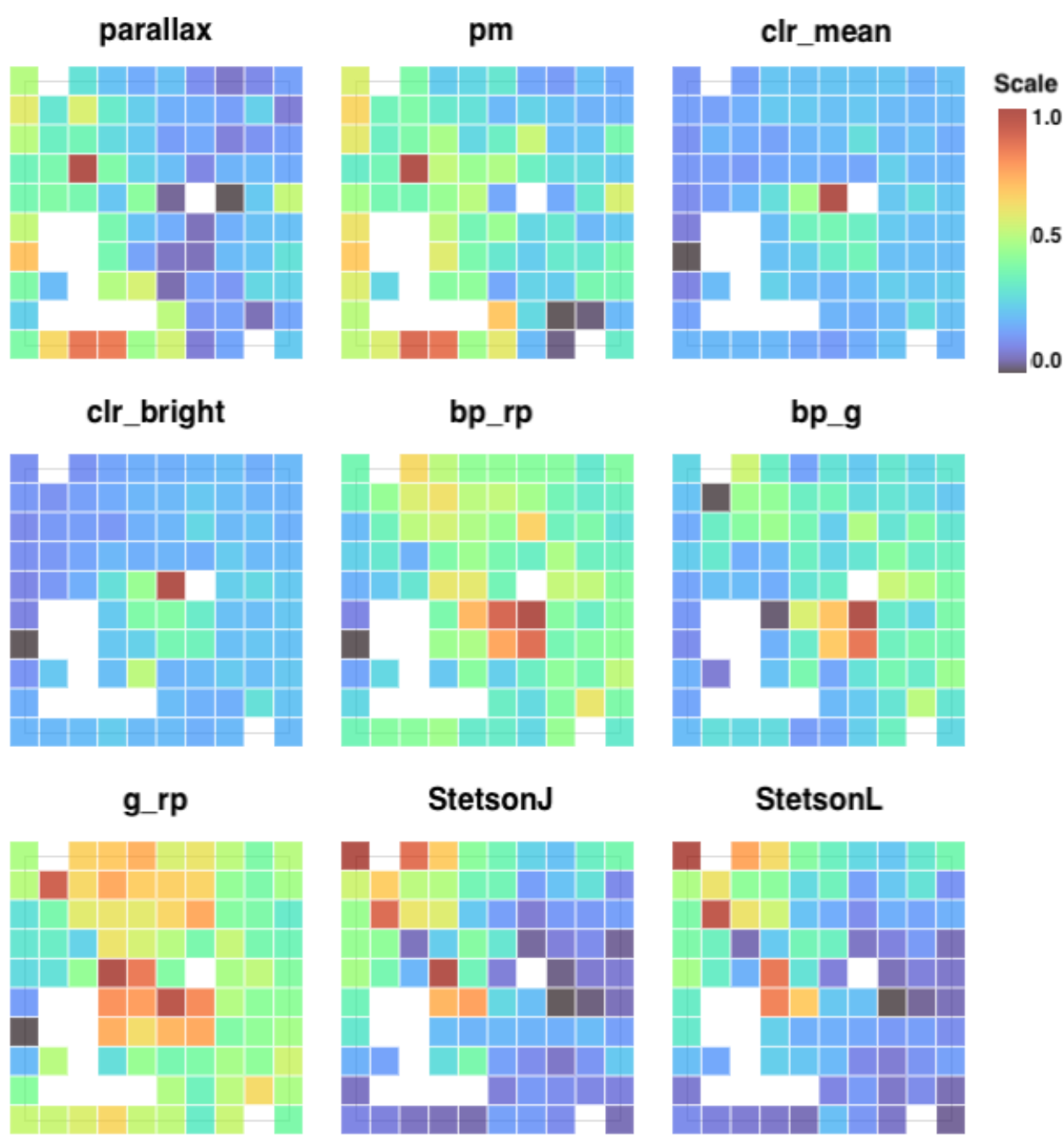}
    \caption{Feature maps for the XGBoost model produced using GTM. Same as for figures \ref{fig:factor_map_g} and \ref{fig:factor_map_r} though for Gaia and colour related features}
    \label{fig:factor_map_clr}
\end{figure}

In Figures \ref{fig:factor_map_g} and \ref{fig:factor_map_r} are a selection of feature maps for features derived from the g and r band light curves. Several further feature maps are shown in Figure \ref{fig:factor_map_clr} representing features derived from a combination of the g and r band light curves and from Gaia DR3. They represent the average feature values of examples assigned to each of the latent space nodes. The feature maps can be used as tools to identify the features most responsible for the assignment of a given class. This is done by comparing class map structures with those within the feature maps. While examination of the feature maps is reserved for the discussion section, it is clear that structures and patterns exists within them that coincide with class map structures. For example, high values for amplitude and variability based features (e.g., \textit{Amplitude}, \textit{Std}, \textit{MedianAbsDev}, and \textit{npeaks}) correspond to outbursting systems; the fewest number of data points, \textit{n\_obs}, are associated with AM CVn, SU UMa and nova classes; and the bluest colours are associated with the AM CVn class.

\subsection{Alert stream pipeline}

With the aim of the alerts filter to minimise the number of possible non-CVs and maximise potential CVs, this was best achieved with the following procedure. The Sherlock contextual classifier was utilised to remove sources within the synonym radius (1.5") of a catalogued active galactic nucleus or nuclear transient. Inspection of light curves of alerting sources (within a 30 day period) removed under these conditions revealed no elimination of known or candidate CVs. To filter out supernova candidates, those sources classified as SN by Sherlock are removed should they meet the following criteria: the closest matching source from the PanSTARRS catalogue (used as the reference source) should have a Star/Galaxy score of less than 0.4 (values range from 0 to 1, where closer to 1 implies a higher likelihood of being a star); and an angular separation from the associated galaxy centre less than the galaxy's semi-major axis size (in arcseconds). Furthermore any source with a Transient Name Server name prefix with 'SN' was also removed. Of the sources remaining with a contextual classification of 'SN', $\sim60\%$ displayed outbursting characteristics where quiescent stages were below the detection limit (likely dwarf novae). The remaining percentage were a mixture of faint sources with no star/galaxy score, several Mira variables, and a classified nova. For removal of variable stars, a simple cross-match with the AAVSO VSX list of Mira variables, Cepheids, RR Lyrae stars (amongst other classes under the variable star umbrella) was performed. Few, if any, of those removed with this variable star filtering method belonged to a member of the confirmed or suspected CV family. With respect to the $\delta m$ criteria, no such filtering is performed in an effort to maximise the number of CVs. It was found that sources with the least amount of variability are assigned the nova-like class, thus a motivating factor in this choice.

Constraining the number of alerts based on several g-r colour metrics, and not just the overall mean, had the desired effect of retaining dwarf nova exhibiting sources. These are outside the epochal or overall mean colour threshold of <= 0.7 during quiescence, but within the threshold during outburst by virtue of the colour measured at their brightest epoch (\textit{clr\_bright}). An approximate quantitative estimate of the effectiveness of this strategy can be given for a month's worth of alerts. For June 2023, 12 confirmed or strong candidate dwarf novae were outside of this threshold based on the mean epochal or overall colour, whereas with the inclusion of the \textit{clr\_bright} quantity only 1 fell outside the threshold.

An additional criteria requiring at least four data points on either the g or r band light curve was also imposed, allowing the majority of features to be derived. Combining all the above criteria, the  number of sources returned per night for input into the ML classifier can be as few as 50, while on other nights over 200 may be available. During June 2023, the filtering output 1283 sources, of which $\sim8\%$ are contained within the Downes Catalog of CVs \citep{RN516} and/or the Ritter Cataclysmic Binaries Catalog v7.24 \citep{RN443}. Approximately 45\% are contained within the AAVSO VSX CV compilation of confirmed or candidate CVs (this includes the Ritter and Downes catalogues). The remainder, those not contained within AAVSO, comprise of: low amplitude slowly varying (month to year long timescales) sources ($\sim30\%$ of the total), a small fraction of which are eclipsing binaries; sources with similar variability to VY Scl and magnetic CVs ($\sim3$ and $4\%$ of the total, respectively); outbursting candidates ($\sim8\%$ of the total); and a combination of sources that have once briefly risen above the limiting magnitude (possible supernovae), and those with too few data points for inference. Further inspection reveals that young stellar objects, candidate AGN, and variable stars provide the majority of contamination. A rough estimated of between 5 and 10\% contamination from these sources is found.

The output of the filter applied to the alerts for June 2023 were fed into the XGBoost classifier with the following findings. The low variability sources are overwhelmingly assigned the nova-like class, while outbursting sources are assigned one of the dwarf nova classes or the AM CVn label. Superoutbursting or candidate superoutbursting systems are largely assigned the SU UMa label with a small amount of mislabelling into the U Gem class. Signatures of Z Cam variability are present within the list of sources assigned this class, while faint blue sources are generally assigned the AM CVn class. As one enters the low sampling regime (fewer than 20 data points) class confusion is evident, though not where outbursting activity is clearly present.

From the June 2023 alerts filter output, we have compiled Table \ref{tab:CV_candidates}. This is a list of candidate CVs we identified that, at the time of writing, are not present in either the Ritter or Downes catalogues, the list of CVs within AAVSO VSX, or within literature as far as we are aware. The prediction of class output by our classifier (along with the class probability) for these candidates is provided. Furthermore, we assign a score based on the strength of their candidacy as members of the CV class. A score of 1 represents a light curve sufficiently sampled for the identification of distinguishing characteristics. Should less well sampled signatures of defining characteristics be present, for example, outbursts not sampled during quiescence, a score of 2 is given. A score of 3 is given to the examples where only faint signatures are present, possibly due to poor sampling.

\begin{table*}
    \caption{New CV candidates identified by our pipeline. Given are the: ZTF object ID; equatorial coordinates at the J2000 epoch; number of suspected dwarf nova outbursts, where (SO) is appended for possible superoutbursts amongst them; g band magnitude range, or r band (appended with r) should insufficient g band data exist (> is prepended should no quiescence brightness be present); light curve duration in days; Gaia BP-RP colour; mean ZTF g-r colour and in brackets the colour at peak brightness, calculated in the manner of the \textit{clr\_mean} and \textit{clr\_bright} features explained in Table \ref{tab:additional_features}; prediction of our classifier; posterior class probability output by our classifier; and the strength of CV candidacy, rated as 1 for the strongest, 3 for the weakest candidates. The table is ordered by class prediction then probability.\label{tab:CV_candidates}}
    \centering
    \begin{tabular}{l|c|c|c|c|c|c|c|c|c|c}
        \hline
        ZTF ID &  R.A. & Dec. & Outb & $\Delta m$ & Dur & BP-RP & g-r & Clf pred & Prob & CV Rating\\
        \hline
        ZTF19aauxfaw & 15:27:39.96 & -19:48:46.17 & 4 & > 17.9-19.1 & 1475 & - & -0.34 (-0.17) & AM CVn & 0.70 & 3\\
        ZTF21aawqeix & 18:49:31.03 & -17:43:54.13 & 4 & > 18.2-19.0 & 810 & - & -0.02 (-0.08) & AM CVn & 0.38 & 2\\
        ZTF18ablpcfv & 19:09:21.11 & -20:01:03.13 & 6-8 & > 17.5-18.7 & 1521 & -0.60 & 0.03 (-0.08) & AM CVn & 0.37 & 3\\
        ZTF23aamdode & 17:08:45.64 & +08:54:51.69 & 1 & > 17.4-20.5 & 44 & - & -0.24 (-0.62) & AM CVn & 0.35 & 3\\
        ZTF19abdmfpn & 17:58:04.69 & +05:28:15.54 & 2 & > 18.9-19.4 & 700 & - & -0.44 (-0.27) & AM CVn & 0.33 & 3\\
        ZTF19aalcaij & 18:01:43.65 & +23:21:11.17 & 4-6 & > 18.9-20.6 & 1409 & - & -0.10 (-0.08) & AM CVn & 0.31 & 2\\
        ZTF19acbwtgi & 22:25:56.91 & +39:26:48.97 & 3 & > 19.3-19.7 & 1375 & - & -0.24 (-0.11) & AM CVn & 0.30 & 3\\
        ZTF18abcysck & 19:03:59.30 & +32:32:37.40 & 12 (SO) & > 18.5-19.7 & 1822 & - & -0.33 (-0.31) & AM CVn & 0.28 & 2\\
        ZTF19aadovsk & 17:44:08.17 & -03:50:46.88 & 5-7 (SO) & > 18.5-19.3 & 1479 & - & -0.16 (-0.04) & AM CVn & 0.26 & 2\\
        ZTF21acbqaqa & 14:50:11.12 & +65:59:42.19 & - & 18.9-20.7 & 654 & - & 0.30 (0.83) & Polar & 1.00 & 2 \\
        ZTF20abpwtmi & 15:38:20.42 & +79:32:26.05 & - & 18.5-20.6 & 1071 & - & 0.38 (0.67) & Polar & 0.96 & 2 \\
        ZTF18abcwxnq & 18:43:26.49 & +06:08:00.90 & - & 17.9-21.7 & 1153 & 1.86 & 0.27 (0.12) & Polar & 0.94 & 2\\
        ZTF18abmrmlu & 23:01:52.75 & +39:50:13.96 & - & 18.7-22.2 & 1791 & 0.91 & 0.41 (-0.12) & Polar & 0.80 & 2 \\
        ZTF18abiklxf & 20:46:40.96 & +22:50:36.20 & - & 17.4-20.3 & 1816 & 1.46 & 0.22 (0.62) & Polar & 0.77 & 2 \\
        ZTF18abnjsqz & 17:40:39.30 & -00:51:46.68 & 2 & > 17.5-19.1 & 547 & - & -0.04 (-0.01) & SU UMa & 0.98 & 2\\
        ZTF18abqbbpq & 17:55:15.36 & +06:57:44.41 & 4 & > 18.6-19.9 & 1501 & - & 0.22 (-0.05) & SU UMa & 0.98 & 3\\
        ZTF19abtnbck & 19:02:38.61 & +26:52:44.76 & 3 & > 18.8-19.7 & 1404 & - & 0.00 (0.00) & SU UMa & 0.98 & 2\\
        ZTF19abdolkk & 19:21:46.43 & -27:54:53.91 & 2 & > 17.8-19.0 & 1454 & - & -0.24 (-0.23) & SU UMa & 0.98 & 2\\
        ZTF19aaprbry & 19:41:32.53 & -07:37:54.12 & 4 & > 18.6-20.0 & 1350 & - & -0.05 (0.06) & SU UMa & 0.98 & 3\\
        ZTF20acufmrl & 02:51:10.20 & +48:39:28.83 & 3 & 18.5-19.9 & 263 & - & -0.07 (-0.06) & SU UMa & 0.97 & 2\\
        ZTF19abjbhmd & 16:55:20.72 & -18:21:58.77 & 5 & > 18.7-19.1 & 1442 & - & -0.35 (-0.29) & SU UMa & 0.97 & 3\\
        ZTF19aalcaij & 18:01:43.65 & +23:21:11.17 & 1 & > 18.9-20.6 & 1409 & - & -0.10 (-0.08) & SU UMa & 0.97 & 3\\
        ZTF19aaxcajp & 21:44:37.10 & +29:30:10.74 & 5 & > 18.3-19.7 & 1499 & - & -0.11 (-0.02) & SU UMa & 0.97 & 2\\
        ZTF19aailtzw & 17:07:44.19 & +02:56:53.04 & 3 & > 18.2-19.6 & 802 & 0.10 & -0.09 (0.02) & SU UMa & 0.94 & 2\\
        ZTF18abcysck & 19:03:59.30 & +32:32:37.40 & 6 & > 18.5-19.7 & 1822 & - & -0.33 (-0.31) & SU UMa & 0.93 & 2\\
        ZTF21aaqwlgv & 18:16:02.45 & +03:07:11.79 & 3 & > 18.3-19.5 & 819 & - & 0.05 (0.11) & SU UMa & 0.92 & 2\\
        ZTF18abklywy & 18:01:53.06 & +04:07:22.51 & 6 & > 18.6-19.9 & 1526 & - & 0.23 (0.08) & SU UMa & 0.91 & 2\\
        ZTF19aadovsk & 17:44:08.17 & -03:50:46.88 & 3 & > 18.5-19.3 & 1479 & - & -0.16 (-0.04) & SU UMa & 0.92 & 2\\
        ZTF18aavtqlz & 17:49:11.47 & +23:58:27.57 & 5 & > 19.2-20.3 & 1265 & - & -0.24 (0.07) & SU UMa & 0.85 & 3\\
        ZTF18abthqde & 19:39:04.33 & +41:53:10.10 & 4 & > 17.4-18.9 & 1760 & - & -0.21 (-0.23) & SU UMa & 0.83 & 2\\
        ZTF20abylzfr & 20:11:08.11 & +84:05:19.21 & 2 & > 17.1-19.7 & 1037 & - & -0.08 (-0.16) & SU UMa & 0.74 & 2\\
        ZTF18absoqce & 23:18:05.90 & +55:58:51.90 & 6 & > 17.9-19.4 & 1773 & - & 0.80 (0.39) & SU UMa & 0.69 & 2\\
        ZTF18ablpcfv & 19:09:21.11 & -20:01:03.13 & 5 & > 17.5-18.7 & 1521 & -0.60 & 0.03 (-0.08) & SU UMa & 0.65 & 3\\
        ZTF19ablvwcu & 20:09:20.00 & +00:22:28.56 & 5 & > 17.7-18.5 & 1331 & - & 0.27 (0.19) & SU UMa & 0.63 & 2\\
        ZTF18abjrekr & 22:00:29.91 & +50:08:47.44 & 5 & > 18.1-19.7 & 1808 & - & 0.20 (0.09) & SU UMa & 0.62 & 2\\
        ZTF18accpsgk & 21:19:34.61 & +38:00:12.90 & 10 & > 17.2-18.0 & 1699 & - & -1.30 (-0.85) & SU UMa & 0.59 & 2\\
        ZTF19ablujxj & 20:36:53.40 & +21:11:06.05 & 7 & > 18.6-20.0 & 1438 & - & -0.03 (0.00) & SU UMa & 0.57 & 2\\
        ZTF18abndsft & 17:25:12.81 & -20:40:48.85 & 4 & 17.7-21.2 & 1474 & 1.69 & 0.74 (0.53) & SU UMa & 0.45 & 2\\
        ZTF18abzmujj & 19:11:51.25 & -05:49:30.43 & 6 & > 18.7-19.6 & 1730 & - & 0.62 (0.41) & U Gem & 0.85 & 1\\
        ZTF18abeajjd & 17:03:58.75 & +15:27:31.78 & 8 & > 18.5-20.7 & 1823 & - & 0.13 (0.18) & U Gem & 0.78 & 1\\
        ZTF19aawxrtk & 18:08:13.30 & +22:51:09.39 & 2 & 16.9-17.2 & 1323 & - & -1.73 (-1.42) & U Gem & 0.68 & 2\\
        ZTF18abloyve & 19:10:41.97 & -26:46:57.55 & 4 & > 16.9-17.9 & 1490 & - & 0.44 (0.20) & U Gem & 0.53 & 2\\
        ZTF18aazeong & 22:24:05.48 & +51:11:42.41 & 10 & 17.3-19.3 & 1847 & 1.15 & 0.20 (0.11) & U Gem & 0.47 & 1\\
        ZTF18abnwfvw & 18:53:33.53 & +22:35:59.41 & 3 & > 16.5-19.9 & 1422 & 1.64 & 0.54 (0.39) & Z Cam & 0.45 & 2\\
        ZTF18abuytrt & 18:13:14.20 & +01:49:02.04 & > 9 & 18.2-20.8 & 1552 & 0.93 & 0.33 (0.40) & Z Cam & 0.35 & 2\\
        ZTF19aarpwtt & 19:54:34.93 & +46:11:08.59 & 10-14 & > 18.8-19.8 & 1485 & - &  0.20 (0.08) & Z Cam & 0.31 & 2\\
        ZTF19ablujxj & 20:36:53.40 & +21:11:06.05 & 12 (SO) & 18.6-20.0 & 1438 & - & -0.03 (0.00) & Z Cam & 0.31 & 2\\
        ZTF18abthqde & 19:39:04.33 & +41:53:10.10 & 5 & > 17.4-18.9 & 1760 & - & -0.21 (-0.23) & Z Cam & 0.30 & 1\\
        ZTF21aaqwlgv & 18:16:02.45 & +03:07:11.79 & 3 & > 18.3-19.5 & 819 & - & 0.05 (0.11) & Z Cam & 0.25 & 2\\
        ZTF18abnjsqz & 17:40:39.30 & -00:51:46.68 & 3 & > 17.5-19.1 & 547 & - &  -0.04 (-0.01) & Z Cam & 0.22 & 2\\
        ZTF19aadospr & 16:53:37.97 & +00:49:11.93 & 4 & > 18.4-19.7 & 805 & 0.36 & -0.04 (-0.07) & Z Cam & 0.21 & 3\\
        \hline
    \end{tabular}
\end{table*}

\section{Discussion} \label{sec:discussion}

\subsection{Classifier performance} \label{sec:discussion_4.1}

The characteristics of the confusion matrix and the blending of class specific structures into one another can be explained in the context of the physical properties of CVs, their evolution, and the properties of their light curves. 


\subsubsection{Class proportions}

A list of thousands of cataclysmic variables accurately labelled into their subtypes based on multi-wavelength photometry with sufficient sampling and spectroscopy for each source is not currently available. While over 15,300 sources have been assigned the CV class according to the AAVSO and BTS, those with ZTF counterparts represented just over 5700 (as of March 2023 when the dataset was constructed). A significant proportion of these belong to the dwarf nova class ($\sim89\%$) of which only 19\% possess labels with the dwarf nova subclass information we required. We were therefore limited to a list of 1439 sources with highly imbalanced class proportions.

Whilst efforts are made to account for this imbalance, the classes lowest in sample size (AM CVn, intermediate polar, and nova) are the weakest performers. Comparisons of light curves associated with each of these classes with remaining classes provide a possible reason for their missclassifications. The Intermediate polar ZTF17aabhicw (see Figure \ref{fig:class_lightcurves}) displays long term variability (weeks to months) as seen in polars, nova-likes and VY Scl (e.g., ZTF18abryuah and ZTF18abmrryp), while ZTF17aabglmw displays occasional dwarf nova outbursts. AM CVns display regular and super outbursts (e.g., ZTF18aaawjmk) and may be faint enough to only be visible during outburst (e.g., ZTF18adkhuxp), overlapping with SU UMa characteristics; longer term changes associated with changed in mass transfer rate (e.g., ZTF18aaabbbv) may also be present. A nova eruption decline (e.g., ZTF19aabjxpe) could be confused with SU UMa systems with long supercycles. 

Despite these issues, the ROC curves and class maps represent a classifier with strong predictive capacity, even for the AM CVn and nova classes. This may be a consequence of features relevant to colour, parallax and proper motion. Nova systems in our sample possess redder colours, while AM CVns typically lie at the blue end of the colour scale. AM CVns are intrinsically faint, thereby are required to be closer than most other CVs to be detectable and induce high values of parallax and, where tangential motion occurs, observable proper motion.

\subsubsection{Dwarf nova classes}

Distinguishing between different classes of dwarf novae primarily hinges on our features' ability to detect the presence of superoutbursts in SU UMa and standstills in Z Cam systems. In a study conducted by \cite{RN508}, an in-depth analysis was undertaken to examine the characteristics of superoutbursts and normal outbursts in dwarf nova systems. The research revealed that Z Cam outbursts typically exhibit a noticeably lower amplitude range, spanning approximately 1-4 magnitudes, compared to the superoutbursts and normal outbursts observed in SU UMa systems, which range from 1-9 and 1-8 magnitudes, respectively. The upper limit for U Gem outbursts falls between these two extremes, with a range of 1-6 magnitudes. Consequently, one would anticipate significantly higher values for amplitude related features for SU UMa compared to the Z Cam systems. Indeed, when examining the g and r band feature maps in Figures \ref{fig:factor_map_g} and \ref{fig:factor_map_r} for amplitude, the difference between the minimum (brightest) and mean or median magnitudes (\textit{dif\_min\_mean} and \textit{dif\_min\_median}), and the number of peaks with amplitudes exceeding 5 magnitudes (\textit{npeaks\_above5}), the highest values are consistently found within the region of GTM latent space occupied by SU UMa systems (see Figure \ref{fig:latent_space_visualisation} class maps). As we shift our focus from the SU UMa region in these class maps to U Gem and then to the Z Cam region, the feature values for the corresponding locations in the feature maps progressively diminish. Our confusion matrix (Figure \ref{fig:confusion_matrix_XGB}), along with those class maps, corroborate with the notion that the most pronounced distinction among dwarf nova subtypes lies between SU UMa and Z Cam.

The semi-regular outbursts in dwarf nova systems exhibit a quasi-periodic pattern when adequately sampled. In ZTF light curves, it is notable that superoutbursts, especially long-lasting ones, tend to receive more comprehensive sampling compared to normal outbursts (refer to Figure \ref{fig:class_lightcurves}). Consequently, the strength or amplitude of signals detected in the Lomb Scargle periodogram can serve as an effective discriminator for distinguishing SU UMa systems from U Gem and Z Cam. Notably, the feature maps within Figures \ref{fig:factor_map_g} and \ref{fig:factor_map_r} illustrate that the amplitude values corresponding to detected frequencies and their harmonics (referred to as \textit{Freq\textbf{i}\_harmonics\_amplitude\_\textbf{j}}; see Table \ref{tab:lc_features}) are consistently higher in regions associated with SU UMa systems than in U Gem and Z Cam associated regions (refer to Figure \ref{fig:latent_space_visualisation} class maps). The peak values of these features are most prominent in regions adjacent to those associated with the AM CVn and nova classes, possibly due to instances where the observational timeline exclusively captures a brightening event, such as a nova eruption or superoutburst.

Figures \ref{fig:factor_map_g} and \ref{fig:factor_map_r} reveal that skewness (\textit{Skew}), standard deviation (Std), and the standstill level (\textit{stdstilllev\_t20s10}), may be used to distinguish Z Cams from other dwarf novae. Our analysis suggests that standstills can significantly influence the magnitude distribution, pushing it towards brighter values. Furthermore, if these standstills persist for an extended period, ranging from weeks to months, they can also reduce the standard deviation, aligning it more closely with that observed in nova-like systems. While regions exhibiting low standard deviation are not exclusive to Z Cam systems, as other dwarf novae with extended periods of quiescence also display this characteristic, what sets Z Cams apart is the normalised brightness within these low standard deviation regions. The standstill level feature aims to pinpoint these distinctive regions within the light curve, effectively distinguishing Z Cam systems from their SU UMa and U Gem counterparts.

When it comes to defining characteristics of U Gem systems, with orbital periods greater than 3 hours, their more massive donor stars and greater mass transfer rates result in accretion disks typically larger than those of SU UMa systems, whose orbital periods mostly lie below 2 hours. Consequently, for the equivalent orbital inclinations U Gem systems have a higher optical quiescent brightness. The combination of ZTF's limiting magnitude and this brightness disparity results in many SU UMa systems only being detected during their outburst phases as opposed to the U Gem class in which quiescence sampling is more likely. This is evident when examining the number of observations \textit{(n\_obs) feature maps in Figures \ref{fig:factor_map_g} and \ref{fig:factor_map_r}, where higher values are present in the U Gem associated region compared to that for SU UMa}.

Expanding upon the topic of intrinsic brightness, sources with lower intrinsic brightness would need to be closer for effective observation, leading to a higher parallax measurement (and possibly proper motion depending on motion in the tangential plane). With the shortest orbital periods of the dwarf nova classes, SU UMa systems are expected to be less luminous, (given equivalent orbital inclinations) for the reasons set out in the previous paragraph, and posses higher parallax values (and proper motion) when compared to their dwarf nova counterparts. These distinctions are indeed evident in the Figure \ref{fig:factor_map_clr} feature maps for \textit{parallax} and \textit{pm}, respectively. Moreover, these arguments align with the observation of fainter absolute magnitudes as well.

The high mass transfer rates characteristic of Z Cam systems drive them to meet the disk instability threshold shortly after a previous outburst. Consequently, during their outburst phases, they tend to spend considerably less time at the minimum brightness level in comparison to other dwarf nova types, as documented by \cite{RN367}. This leads to recurrence periods typically falling within the range of 10 to 30 days, exemplified by systems like ZTF17aaaeepz. It is reasonable to anticipate that the outburst recurrence period, a parameter that the Lomb Scargle periodogram's maximum power frequency (\textit{freq\_pwr\_max}) aims to characterise, could offer some level of discrimination between Z Cam systems and their dwarf nova counterparts.

However, upon scrutinising the corresponding feature maps for \textit{freq\_pwr\_max} (within Figures \ref{fig:factor_map_g} and \ref{fig:factor_map_r}), it becomes evident that distinguishing between these types is challenging. For potential insights into this challenge, one may refer to the findings of \cite{RN508}. Notably, while the average recurrence periods for the U Gem class tend to be longer than those of Z Cam systems, in excess of 50 days, there is an overlapping range with Z Cam recurrence periods. This overlap is also observed in the case of the SU UMa class, where recurrence periods span from 3 to 300 days. Additionally, factors such as the presence of extended standstills in Z Cam systems (e.g., ZTF17aabunpt; Figure \ref{fig:class_lightcurves}) and the limited sampling of normal outbursts contribute to the complexity of estimating this type of periodicity.

An examination of light curves for systems that fall between the latent space nodes associated with U Gem and Z Cam classes (see Figure \ref{fig:latent_space_visualisation}) further confirms this recurrence period overlap, as does the overlap between the SU UMa and U Gem classes. Within this continuum also lie the rapidly outbursting SU UMa subtypes, ER UMa, underscoring the significance of recurrence period overlap as a primary contributor to the confusion among dwarf nova subclasses.

\subsubsection{AM CVn}

For the remainder of Section \ref{sec:discussion_4.1}, in order to facilitate our discussion and interpretation of the class and feature maps, we may refer to specific nodes (squares) by a simple coordinate system (x, y). The value of x denotes the square number (1-10) from left to right, while the value of y signifies the square number (1-10) from bottom to top.

As previously discussed in the introduction, AM CVn systems tend to be bluer than their hydrogen-rich CV counterparts and are generally of lower luminosity. While superoutbursts are observed in AM CVn systems \citep{RN476}, they tend to be of shorter duration, typically lasting 5-6 days, and display lower amplitude (4-6 magnitudes) in contrast to superoutbursts in SU UMa systems, which often extend beyond 10 days and can, in the case of the WZ Sge subclass of SU UMa, reach amplitudes exceeding 6 magnitudes. Additionally, normal outbursts have also been observed in AM CVn systems, occurring on the fading tail of superoutbursts \citep{RN433}.

Upon scrutiny of feature maps, it becomes apparent that features such as the mean, median, minimum, and maximum magnitude derived from g-band light curves (Figure \ref{fig:factor_map_g} feature maps) do not strongly differentiate AM CVn systems from other classes, contrary to the expectation of higher (and consequently fainter) values. Similar observations hold true for the r-band (Figure \ref{fig:factor_map_r}), with the exception of the minimum magnitude in the r-band (\textit{min\_mag\_r}), where notably elevated (i.e., fainter) values cluster around node (1,4), associated with the highest AM CVn probability (see Figure \ref{fig:latent_space_visualisation} class maps). One possible explanation for these findings is that accretion discs in AM CVns are smaller than those in hydrogen CVs, truncated by the smaller Roche lobe geometry. As emissions in the r-band primarily originate from the cooler outer regions of the accretion disc, the effective surface area of these regions is considerably smaller for the compact AM CVn discs.

To become detectable, AM CVn systems would be required to be situated at closer distances, thereby inducing higher parallax measurements and, in cases where tangential motion occurs, observable proper motion (\textit{pm}). While node (1,4) within the corresponding feature maps in Figure \ref{fig:factor_map_clr} may not contain the highest values (which are located at node (3,7) and associated with the SU UMa region), they still exhibit values sufficiently high enough to align with our expectations when compared to regions associated with other classes.

The average ZTF g-r colours, along with Gaia colours (involving RP data), are strong discriminators effectively separating AM CVn systems from other classes, as evident in Figure \ref{fig:factor_map_clr}. However, when it comes to outburst-specific features (e.g., \textit{npeaks\_2to5}; Figures \ref{fig:factor_map_g} and \ref{fig:factor_map_r}), their effectiveness diminishes. Contributing factors to this reduced performance may be due to the scarcity of AM CVn examples within the dataset, coupled with variations in observational time spans and the sampling of their light curves. Consequently, this diversity results in a variety of light curve profiles, as depicted in Figure \ref{fig:class_lightcurves}, where the number of sampled outbursts range from several to none at all. An examination of sources projected onto latent space regions where the boundaries between AM CVn and SU UMa classes, as well as between AM CVn and nova classes, blend together (see Figure \ref{fig:latent_space_visualisation}), suggests that these factors contribute significantly to the observed classification ambiguity.

\subsubsection{Novae}

Despite a low sample size, the nova class achieves a recall score of 0.50 and a precision of 0.64. A significant source of false-positive predictions in the nova class can be attributed to the SU UMa class. A possible explanation could simply be due to nova dataset examples consisting largely of extragalactic sources, visible during the time of peak eruption brightness. These light curves bear a resemblance to those of SU UMa systems where only one outburst (often a superoutburst) has been sampled. Consequently a low number of observations is associated with the class, as is the case for SU UMa systems.

Two members of the nova class within the test set have missclassifications as VY Scl. A possible explanation could be provided by ZTF21abmbzax (example light curve in Figure \ref{fig:class_lightcurves}), which displays a 'dust dip' explained as being generated by dust in the eruption ejecta absorbing photons and re-emitting in the infra-red \citep{RN180}. This characteristic resembles a VY Scl low state excursion. Another eruption light curve profile mentioned in \cite{RN180} is that which exhibits a 'flat top and jitters' - cuspy profiles at eruption maximum. This is seen in ZTF19abirmkt, and could be responsible for missclassifications of novae as magnetic CV members. Projections of these sources onto the GTM latent space of Figure \ref{fig:latent_space_visualisation} align with these interpretations, with ZTF21abmbzax projected onto node (6,3), located in between the nova and VY Scl structure cores, and ZTF19abirmkt projected onto node (3,3) located between the nova and polar structure cores.

\subsubsection{Remaining classes}

The separation between the intermediate polars, polars, nova-likes, and the VY Scl nova-like subtype arises from several physical properties manifested in their light curves, as discussed in the introduction. As just demonstrated in previous subsections, comparison of the g and r band feature maps within Figures \ref{fig:factor_map_g} and \ref{fig:factor_map_r} with the class probabilities depicted in the Figure \ref{fig:latent_space_visualisation} class maps, help highlight the light curve attributes most relevant for class separation.

The VY Scl class stands out with its deep low brightness state excursions such that low values of \textit{eta\_e} appear in the relevant g and r band feature maps near node (7,1), associated with the highest VY Scl class probability (see class maps). This feature reflects the degree of independence between successive data points, where magnetic systems exhibit higher values due to hourly timescale variations, while VY Scl systems show lower values due to longer timescale variations. Furthermore, VY Scl low state excursions can induce a high skewness in magnitudes (\textit{Skew}), and due stable and prolonged high-brightness states, give rise to the highest standstill level values (\textit{stdstilllev\_t20s10}), as feature and class maps demonstrate. 

Eclipses within the nova-like class, as exemplified by ZTF18abajshu in Figure \ref{fig:class_lightcurves}, push the standstill level into a range occupied by Z Cams, potentially causing confusion between these two classes. Confusion also arises between nova-likes and the SU UMa class. The light curves of sources where such confusion occurs are marked by a limited number of data points, this is seen in the \textit{n\_obs} feature maps for nodes (6,6) and (6,7), situated where the associated class structures are closest together. Based on the current feature set, our model finds difficulty in distinguishing systems visible only during outbursts from nova-likes with limited observational epochs, though overall, nova-likes remain distinguishable from the other classes.

The lowest standard deviation (\textit{Std}) and absolute median deviation (\textit{MedianAbsDev}) values are associated with the intermediate polar and nova-like classes, as seen in the feature maps. This aligns with the less frequent low state excursions observed in intermediate polars and nova-likes compared to polars and VY Scl systems in the ZTF light curves.

As explained by \cite{RN431}, most intermediate polars possess accretion disks truncated at inner radii due to the white dwarf's magnetism. This may lead to dwarf nova outbursts characterised by lower amplitudes and shorter durations. The mixture of outbursting and non-outbursting intermediate polars, coupled with less distinct outburst profiles, contributes to feature maps displaying lower amplitude and variability-related values for intermediate polars compared to dwarf novae. Non-outbursting intermediate polars may provide an explanation for confusion with polars, indeed, this is supported by the projection of intermediate polar ZTF18abaiuvj (Figure \ref{fig:class_lightcurves}) onto a region associated with polars within the GTM latent space (Figure \ref{fig:latent_space_visualisation}).

\subsubsection{Evolutionary factors}

Separating cataclysmic variables into distinct classes is one that is a challenge for experts on the subject who must wrestle with the fact that as these systems evolve, they transition from displaying traits characteristic of one class to another such that boundaries between classes are blurred (e.g., \citealt{RN9,RN445,RN421,RN495,RN442}). This is a consequence of the requirement for stable mass transfer that depends upon the magnetic braking and gravitational wave radiation angular momentum loss mechanisms to drive systems to shorter periods. In so doing this allows the mass losing donor to continue transferring mass by maintaining contact with its Roche lobe (e.g., \citealt{RN155, RN445}).

The shortening of orbital periods, donor composition changes, and shrinkage of the accretion disk amongst several other factors drive the class transitions. At long periods (typically 3-6 hours) the high mass transfer rates allow the accretion disk to be maintained in a stable hot viscous state, such that no dwarf nova outbursts are observed, these systems form the nova-likes. As the mass transfer rates drop, the accretion disk straddles the stability threshold, below which the disk is cool, non-viscous and unstable to dwarf nova outbursts \citep{RN442}. Systems lying close to this threshold form the Z Cams, with periods of standstill, akin to nova-likes, and outbursting episodes typical of dwarf novae. The continuing evolution induces an unstable disk, where we see a transition to the semi-regular outbursts of U Gems and then the SU UMa systems where the much shorter periods (< 2 hours) introduce tidal effects driving the superoutbursts they are known for. SU UMa systems comprise ER UMa subtypes, where above average mass transfer rates lead to short superoutburst recurrence periods and rapid fire normal outbursts that may cause confusion with the Z Cam class; and the WZ Sge subtypes (shortest period SU UMa) whose donor composition now drives an increase in orbital period. 

As with hydrogen CVs, helium CVs (AM CVns) undergo evolution. Due to a degenerate/semi-degenerate donor, evolution once mass transfer starts (at periods of 5-10 minutes) is towards longer periods, during which accretion may transition from direct (no disk), to hot stable, then unstable disks subject to the He CV equivalent of dwarf nova outbursts \citep{RN315,RN70}. 

Nova eruptions are a possibility for all systems should conditions for hydrogen fusion be present under degenerate conditions on the WD surface; this is far more likely to occur for the highest mass transfer rate systems with high mass WD accretors (e.g., \citealt{RN428,RN167,RN96,RN62}. 

The presence of strong magnetic fields for the intermediate polar or polar label requires observations of pulsed X-rays and/or polarimetry to complicate matters further. In addition to the above, one must factor in the orbital inclination that determines the overall emission contribution from the accretion disk and thereby impacts measurements such as colour and brightness.

The evolutionary changes are evident in many of our light curves. CR Boo (ZTF18adkhuxp; Figure \ref{fig:class_lightcurves}), is an AM CVn with a standstill to its name \citep{RN511}; high mass transfer rate systems residing amongst the U Gem class manifest as dwarf nova outbursts with very short recurrence times indicative of a Z Cam class; the ER UMa subclass \citep{RN529} of SU UMa systems may also be confused with the Z Cam class due to their high mass transfer rates and rapid outbursts. With respect to intermediate polars a range of light curve morphologies are possible (e.g., \citealt{RN565}). Short duration low state transitions, dwarf nova outbursts and more stable long term light curves are present within our light curve sample, consequently, a confusion with any of the other classes is possible. Constructing a classifier in light of these intricacies will naturally produce class confusion despite incorporating a wide ranging feature set inclusive of astrometric data and an attempt to produce a dataset with accurate class labels. The confusion matrix and class maps in combination with the example light curves displayed in figure \ref{fig:class_lightcurves} are a visual representation of this very aspect of CV classification.

\subsection{Pipeline implementation}

A substantial portion of the alert stream filter consists of either known or candidate CVs (according to the AAVSO VSX list). This is positive news, indicating that the filter effectively retains them in the stream. Consequently, the undiscovered CV candidates have a promising likelihood of being contained within the remaining alerts that have successfully passed through the filter. The approximate class proportions of the confirmed or candidate objects are as follows: 20\% SU UMa (including the WZ Sge and ER UMa subtypes), 4\% Z Cam, 3\% U Gem, 59\% dwarf novae without further subdivision, 3\% magnetic CVs, 6\% nova-likes (including subclasses), and less than 1\% AM CVn. The remaining confirmed or candidate CVs form a mixture of several sources labelled as novae due to an eruption that may have occurred before ZTF observations, a recurrent nova, and CVs without further subdivision. These proportions stem from a variety of factors that may include: the frequent occurrence of alert-triggering events in dwarf novae leading to their relatively higher representation; the inherent faintness of short (or ultrashort) period CVs, making their detection less probable; the need for supporting evidence, such as periodic variability on short timescales (minutes to hours), polarimetry, and/or X-ray emission, to confidently confirm a CV as magnetic; and the establishment of specific thresholds in our alert filtering, e.g., excluding CVs with a g-r colour index exceeding 0.7.

The substantial contribution of low-variability sources among the remaining filter targets results from the omission of a magnitude change condition. Nevertheless, it was observed that incorporating such a condition restricted the detection of confirmed/candidate outbursting CVs, unless a considerably low threshold was applied. Notably, our classifier overwhelmingly assigns the nova-like label to low (or slowly varying) sources, thereby enabling the classifier to allocate the remaining higher variability sources into distinct classes. Nonetheless, we retain the option to implement a magnitude change criteria should we choose to focus on specific variability types.

Referring to Figure \ref{fig:cmd_gminusr}, configuring the filter to retain alerting sources with a ZTF g-r colour of <= 0.7 is expected to encompass the vast majority of the shortest period systems, SU UMa, and AM CVn candidates, along with a significant portion of the remaining classes. However, expanding the filter to include all examples would inevitably lead to a rise in contamination from non-CVs, such as Mira variables and AGN candidates (as observed in the June 2023 sample). Similar to the magnitude change filtering, we are actively exploring the option to adjust the colour constraint, aiming to focus on specific CV subclasses.

The ML classifier demonstrates its greatest strength when applied to the filter output by effectively distinguishing between outbursting and non-outbursting sources, a characteristic mirrored in the test set predictions. Also mirroring the test set results is the further separation of confirmed, candidate, or likely (from our inspection) SU UMa from Z Cam sources; and the separation of light curves with polar and VY Scl like variability assigned to those respective classes. However, when we enter the low sampling regime, the classifier struggles to assign alerting sources into what we would consider the appropriate class. For example several poorly sampled though likely outbursting systems (where quiescent magnitudes are not sampled) are assigned the nova-like or polar classes. However, on the whole, these sources tend to be assigned one the dwarf nova classes or the AM CVn class (should an especially blue colour be calculated).

\section{Conclusions}

In this paper, we developed and applied a machine learning pipeline to detect and categorise cataclysmic variables (CVs) and their subtypes from the ZTF alerts stream. Our pipeline's alert filtering stage effectively retains both known and potential CVs across various subclasses, thanks to a multi-parameter g-r colour threshold and the omission of a magnitude change condition. This approach accommodates colour changes during dwarf nova outbursts.

The performance of our ML classifier is largely dependent on the ability of our dataset to provide an accurate representation of the diversity within the CV population. This diversity is clearly present in the example light curves (see Figure \ref{fig:class_lightcurves}), however, imbalance in this diversity (class imbalance) and commonalities in the types of photometric variability between classes renders CV subtype classification a particularly challenging task. Evolutionary factors drive the difficulty in arriving at concrete class labels both for experts in the subject and our ML classifier. The challenge is compounded by inadequate sampling of light curves. Despite these difficulties, an exhaustive examination of several ML algorithms, trained with a comprehensive feature set, and operating under a selection of class balancing and feature selection techniques, yielded a classifier with a prediction pattern that can be understood in the context of CV evolution.

Latent space representations of this prediction pattern using GTM (class maps) provide an easily interpretable avenue for visualising this evolution. The accompanying feature maps provide a convenient method of finding those features most relevant for a model's assignment of a given class. They also provide us with the properties that contribute to classification error, where in many cases the answers are linked to evolutionary factors. Though not explored in this work, these feature maps provide a method to pare down the feature set by eliminating features that provide little benefit for discrimination between classes.

Implementation of the pipeline on the ZTF stream has, over the period of June 2023 alone, yielded a sample of new CV candidates, These are largely of an outbursting nature, with several magnetic CV candidates. With further improvements to the pipeline underway, such as filter threshold adjustments and inclusion of computer vision techniques to provide an automated interpretation of salient light curve characteristics, we aim to reduce contamination of non-CVs (e.g, Mira variables and active galactic nuclei) and produce a ML classifier with greater class distinction powers. 

Given the fuzzy boundary between CV subclasses for the reasons mentioned, it may be prudent to apply stricter criteria for dataset inclusion, focusing only on clear examples of a given class. With this approach, one relies less on definitive class labels, but more on the probability of class belonging. Alternative approaches may include adopting a multi-label approach that takes into consideration class boundary crossing variability, or an unsupervised learning strategy that does away with existing class labels, tasking algorithms with finding similarities, differences and structure in the data itself.

\section*{Acknowledgements}

D. Mistry acknowledges a PhD studentship from Liverpool John Moores University (LJMU) Faculty of Engineering and Technology. M.J. Darnley receives funding from UK Research and Innovation (UKRI) grant number ST/S505559/1. C.M. Copperwheat receives funding from UKRI grant numbers ST/X005933/1 and
ST/W001934/1.

This research has made use of the International Variable Star Index (VSX) database, operated at AAVSO, Cambridge, Massachusetts, USA.

We would like to thank the anonymous referee for taking the necessary time and effort to review the manuscript. We sincerely appreciate all your valuable comments and suggestions, which helped us in improving the quality of the manuscript.

\section*{Data Availability}

The list of cataclysmic variable sources used to generate the machine learning dataset are available as online supplementary material.



\bibliographystyle{mnras}
\bibliography{manuscript.bib} 




\appendix



\bsp	
\label{lastpage}
\end{document}